\documentclass[aps,prx,10pt,amsmath,amssymb,floatfix,notitlepage,twocolumn]{revtex4-2}

\usepackage[colorlinks=true,linkcolor=blue]{hyperref}
\usepackage{graphicx}
\usepackage{times}
\usepackage{amssymb}
\usepackage{amsmath}
\usepackage{mathtools}
\usepackage{amsthm}

\usepackage{bm}         
\usepackage{color}
\usepackage[utf8]{inputenc}
\usepackage[T1]{fontenc}
\usepackage[normalem]{ulem}

\begin{document}

\title{Quantum dot energy levels in bilayer graphene: Exact and approximate study}

\author{G. Giavaras}
\affiliation{Instituto de Ciencia de Materiales de Madrid (ICMM),
Consejo Superior de Investigaciones Cient\'ificas (CSIC), Sor
Juana In\'es de la Cruz 3, 28049 Madrid, Spain}

\begin{abstract}
In bilayer graphene the exact energy levels of quantum dots can be
derived from the four-component continuum Hamiltonian. Here, we
study the quantum dot energy levels with approximate equations and
compare them with the exact levels. The starting point of our
approach is the four-component continuum model and the quantum dot
is defined by a continuous potential well in a uniform magnetic
field. Using some simple arguments we demonstrate realistic
regimes where approximate quantum dot equations can be derived.
Interestingly these approximate equations can be solved
semi-analytically, in the same context as a single-component
Schr\"odinger equation. The approximate equations provide valuable
insight into the physics with minimal numerical effort compared
with the four-component quantum dot model. We show that the
approximate quantum dot energy levels agree very well with the
exact levels in a broad range of parameters and find realistic
regimes where the relative error is vanishingly small.
\end{abstract}

\maketitle

\section{Introduction}

The versatile electronic properties of bilayer graphene
(BLG)~\cite{rozhkov2, cann}, along with experimental advancements
have enabled the fabrication of highly controllable quantum
devices which might find applications in emerging quantum
technologies. Quantum dots (QDs) are among these quantum devices
and various studies have explored control and manipulation of
confined charge, valley, and spin states~\cite{gachter, rakhm,
costa15, garreis, bucko, costa14, pereira, moller, mirzakhani,
eich18, knothe1, knothe2}. In BLG confined quantum states can be
defined by electrostatic gates providing a simple and attractive
platform to realize tunable quantum dots~\cite{eich18, recher}. A
perpendicular magnetic field can lift the valley degeneracy
offering additional control of the discrete energy levels and
quantum states of the QD. The resulting QD can be used to host
qubits~\cite{recher, bucko} which can be efficiently controlled
with external electric and magnetic fields. Beyond the
spectroscopic investigation and characterization of single QDs in
BLG~\cite{moller}, coherent charge oscillations have been
successfully demonstrated~\cite{hecker} and measurements of
valley~\cite{garreis} and spin lifetimes~\cite{gachter} have been
performed.

From a theoretical point of view, many theoretical works on QDs in
BLG focus on hard-wall potential wells because the resulting QD
equations are drastically simplified and can be studied
semi-analytically~\cite{recher, bucko, pereira}. However,
hard-wall potentials are usually not realistic enough and
therefore fail to capture the complete physics. In addition, the
majority of theoretical works uses solely numerical routines to
study the energy levels of the QDs, without providing any
essential pedagogical insight into the underlying physics. This
occurs despite the fact that many interesting physical regimes in
a bilayer QD system can be treated approximately and with high
accuracy. Approximate methods are therefore a necessary and
powerful tool~\cite{giavaras10, giavaras21} to better understand
the properties of QDs in BLG.

In the present work we start with the four-component continuum
model and define the QD with a continuous potential well in a
uniform magnetic field. We explore potential wells with soft-wall
and hard-wall profiles and demonstrate that it is possible to
derive approximate single-component equations to calculate the
energy levels of the QD in a broad range of parameters. The main
advantage is that the approximate equations can be solved
semi-analytically in the same context as a single-component
Schr\"odinger equation. We can thus obtain valuable insight into
the physics and predict the correct features of the QD energy
levels at different magnetic fields and potential well profiles.
We show that the approximate QD levels are in a very good
agreement with the exact QD levels derived from the four-component
continuum model. Interestingly, in some realistic QD regimes the
induced errors are vanishingly small indicating that the
approximate equations are reliable and capture all the important
physics.

We describe in Sec.~\ref{model} the physical model of the QD in
BLG based on the four-component continuum model. We continue in
Sec.~\ref{first} to detail the necessary steps that allow us to
derive the approximate QD equations. In Sec.~\ref{cases} we
identify with specific examples some QD regimes that can be
studied approximately. In Sec.~\ref{compa} a detailed comparison
is made between the exact and the approximate QD energies
demonstrating the accuracy of our method. The conclusions of the
work are summarized in Sec.~\ref{conclu} and some further results
are presented in two Appendixes.

\section{Physical model}\label{model}

In this section we describe the quantum dot formed in a bilayer
graphene lattice with Bernal stacking~\cite{rozhkov2}. In the
four-component continuum model the envelope functions describing
the QD system in the vicinity of the $K$ valley can be written in
the form: $\Psi=(\psi_{A}, \psi_{B}, \psi_{B'}, \psi_{A'})^{T}$,
where the subscripts $A$, $B$ and $A'$, $B'$ denote the
corresponding sublattices in the upper and lower graphene layers
respectively. When trigonal warping effects are ignored the QD can
be accurately described by the Hamiltonian
\begin{equation}\label{Hmatrix}
H_{\rm QD} = \left(\begin{array}{cccc}
  V_1 & u_0\pi & t_c & 0 \\
  u_0\pi^{\dagger} & V_1 & 0 & 0 \\
  t_c & 0 & V_2 & u_0\pi^{\dagger} \\
  0 & 0 & u_0\pi & V_2 \\
\end{array}\right),
\end{equation}
with $\pi = (p_x + i p_y)$, $\pi^{\dagger} = (p_x - i p_y)$, and
$(p_x, p_y)={\bf p}$ being the 2D momentum operator. We consider
an external magnetic field $B$ perpendicular to the BLG
($z$-direction) described by the vector potential ${\bf A}$. For a
cylindrically symmetric system the vector potential is nonzero in
the azimuthal $\theta$-direction, ${\bf A}= (0, A(r), 0)$, with
$A(r)=B r/2$ and the operator ${\bf p}$ is replaced by ${\bf
p}+e{\bf A}$ when $B\ne0$. The parameter $t_c$ is a constant
interlayer coupling which is taken to be $t_c=400$ meV together
with the lattice parameter $u_0 = 10^{6}$ m s$^{-1}$.

We also add to the bilayer system an electrostatic potential term,
$V_{\rm QD}(r)$, that models the QD potential well, and
specifically we consider that
\begin{equation}\label{Vterms}
V_{1,2}(r) = V_{\rm QD}(r) \pm \tau \frac{V_{\rm b}(r)}{2},
\end{equation}
where $V_{\rm b}$ denotes a bias between the two graphene layers.
In a QD system $V_{\rm b}$ is usually assumed to be constant,
however, the approximate method developed in Sec.~\ref{first} is
general enough and can be applied to QDs with a position dependent
bias. We assume a gate-induced QD and model the QD potential well
as
\begin{equation}\label{potential}
V_{\rm QD}(r) = \frac{-V_W}{ \cosh\left( r/L_{W} \right)^{s}}.
\end{equation}
The exact profile of the QD potential is determined by the
parameters $V_W$, $L_W$, $s$. We consider a few values of $s$ in
order to simulate either a smooth potential well (soft-wall) which
arises for smaller $s$ values, or a much steeper potential well
(hard-wall) with a relatively flat bottom. In Eq.~(\ref{Vterms})
we take $\tau=1$ for the $K$ valley and also consider $\tau=-1$
for the $K^{'}$ valley. Note, however, for $K^{'}$ the two
graphene layers in the envelope functions are exchanged.

Because both $A$ and $V_{\rm QD}$ depend on the radial only
coordinate the resulting eigenvalue problem describing the QD,
$H_{\rm QD}\Psi(r,\theta) = E\Psi(r,\theta)$, can be simplified by
eliminating the $\theta$ dependence. This is achieved with the
substitution
\begin{equation}
\Psi(r, \theta) =\frac{1}{\sqrt{r}}\left(\begin{matrix}
\phi_{A}(r)e^{i m \theta}
\\
i \phi_{B}(r)e^{i (m-1)\theta}
\\
\phi_{B'}(r)e^{i m \theta}
\\
i \phi_{A'}(r)e^{i (m+1) \theta}
\end{matrix}\right),
\end{equation}
which leads to the following coupled equations for the radial
components of the QD states
\begin{subequations}
\begin{eqnarray}
\left(\gamma \frac{d}{dr}+L_1\right)\phi_{B} +t_c\phi_{B'}
&=&(E-V_1)\phi_{A}, \label{dota}\\
\left(-\gamma \frac{d}{dr}+L_1\right)\phi_{A} &=& (E-V_1)\phi_{B}, \label{dotb}\\
\left(\gamma \frac{d}{dr}+L_2\right)\phi_{A'} +t_c\phi_{A}
&=& (E-V_2)\phi_{B'}, \label{dotc}\\
\left(-\gamma \frac{d}{dr}+L_2\right)\phi_{B'}  &=&
(E-V_2)\phi_{A'},\label{dotd}
\end{eqnarray}
\end{subequations}
with the parameter $\gamma = u_0 \hbar$ and the terms
\begin{eqnarray}
L_{1,2}(r)=\mp \gamma\frac{m \mp \frac{1}{2}}{r} \mp
\gamma\frac{e}{\hbar}A(r),
\end{eqnarray}
which account for the angular momentum, denoted by the integer
$m$, and the magnetic vector potential in the azimuthal direction.

\section{Derivation of approximate quantum dot equations}\label{first}

In this section we outline the basic steps that allow us to
simplify the four-component continuum model of the QD and derive
an approximate single-component equation. This equation gives the
QD energies of the confined states. Alternatively, we could use as
a starting point the well known two-component continuum model and
perform further simplifications in order to obtain an approximate
equation. In this approach, however, the approximate energies can
be computed only within the range of validity of the two-component
model. For this reason, this approach is expected to be less
accurate and for clarity of presentation we start directly with
the four-component model.

Analytical calculations in the Landau regime~\cite{peeters} have
shown that the two components $\phi_{A}$ and $\phi_{B'}$ can be
expressed with the same analytical functions and the difference
$|\phi_{A}|-|\phi_{B'}|$ becomes smaller as $m$ increases. Based
on these observations the first step in our method is to eliminate
the two components $\phi_{B}$ and $\phi_{A'}$ and derive the
coupled equations for the components $\phi_{A}$ and $\phi_{B'}$.
We use Eqs.~(\ref{dota}) and (\ref{dotb}) to eliminate $\phi_{B}$,
and Eqs.~(\ref{dotc}) and (\ref{dotd}) to eliminate $\phi_{A'}$. A
simple analysis leads to
\begin{subequations}
\begin{eqnarray}
\left( \frac{d^2}{dr^2}+a_1\frac{d}{dr} + b_1   \right)\phi_{A} =
\frac{t_c}{\gamma^2}(E-V_1) \phi_{B'}, \label{pA} \\
\left( \frac{d^2}{dr^2}+a_2\frac{d}{dr} + b_2   \right)\phi_{B'}
=\frac{t_c}{\gamma^2}(E-V_2) \phi_{A}, \label{pB}
\end{eqnarray}
\end{subequations}
with the coefficients
\begin{subequations}
\begin{eqnarray}
a_j(r)  &=& \frac{ V^{'}_{j} }{E-V_j},  \\
b_j(r)  &=& \frac{L_j}{\gamma} \frac{V^{'}_{j}}{(E-V_j)} +
\frac{L^{'}_{j}}{\gamma} - \frac{L^{2}_{j}}{\gamma^2} +
\frac{(E-V_j)^2}{\gamma^2},
\end{eqnarray}
\end{subequations}
$j=1,$ 2 and prime denotes the first derivative with respect to
$r$. We then perform the transformation
\begin{equation}
\left(\begin{array}{c}
 \phi_{A} \\
 \phi_{B'}\\
\end{array}\right) = \left(\begin{array}{cc}
   (E-V_1)^{1/2}  & 0 \\
    0  &  (E-V_2)^{1/2} \\
\end{array}\right)
\left(\begin{array}{c}
  w \\
  \nu \\
\end{array}\right),
\end{equation}
and derive the two coupled equations for the functions $w$ and
$\nu$
\begin{subequations}
\begin{eqnarray}
\left(\frac{d^2}{dr^2} + q_{1} \right)w   = \frac{t_c}{\gamma^2}[(E-V_1)(E-V_2)]^{1/2}  \nu, \label{w}\\
\left(\frac{d^2}{dr^2} + q_{2} \right)\nu =
\frac{t_c}{\gamma^2}[(E-V_1)(E-V_2)]^{1/2} w, \label{n}
\end{eqnarray}
\end{subequations}
with
\begin{equation}
q_j(r) = b_j - \frac{1}{2}\frac{V^{''}_{j}}{(E-V_j)} -\frac{3}{4}
\left(\frac{ V^{'}_{j} }{ E-V_j } \right)^2,
\end{equation}
Equations~(\ref{w}) and (\ref{n}) are exact and to proceed with
some approximations we set
\begin{equation}\label{lambda}
\frac{1}{ \lambda } = \frac {\nu} {w},
\end{equation}
and assume the function $\lambda$ to be approximately constant in
the region where the QD states are confined. This regime has been
shown to be valid for the Landau states~\cite{peeters} but we
expect deviations from this regime when $V_W\ne0$, since the
resulting confining region of the QD states is now modified.
However, these deviations are small when the character (profile)
of the Landau states is not significantly altered due to
$V_W$~\cite{giavaras21}, therefore $\lambda$ can be assumed to be
approximately constant beyond the Landau regime. This case has
been demonstrated for different potential wells defining QDs in
monolayer graphene~\cite{giavaras21, giavaras12, giav};
consequently, a similar framework should be valid in bilayer
graphene. A characteristic example of small deviations from the
Landau regime due to $V_W\ne0$ involves the creation of an extra
node in the induced QD states~\cite{giavaras22}. Provided this
node lies in a region where the amplitude of the states is
exponentially small it can only weakly affect the original
character of the Landau states~\cite{giavaras22}. In
Appendix~\ref{wkb} we further examine the condition $w=\lambda
\nu$.

Substituting Eq.~(\ref{lambda}) into Eqs.~(\ref{w}) and (\ref{n})
and neglecting terms containing derivatives of $\lambda$ we derive
from Eq.~(\ref{w}) that
\begin{equation}\label{l1}
\frac{1}{ \lambda } = \frac{\gamma^2}{ t_c \sqrt{\kappa} }
\left(\frac{1}{\nu} \frac{d^2\nu }{dr^2} + q_1\right),
\end{equation}
and simultaneously Eq.~(\ref{n}) gives
\begin{equation}\label{l2}
\lambda = \frac{\gamma^2}{t_c \sqrt{\kappa} }
\left(\frac{1}{\nu} \frac{d^2\nu }{dr^2} + q_2\right).
\end{equation}
with $\kappa=(E-V_1)(E-V_2)$. Combining Eqs.~(\ref{l1}) and
(\ref{l2}) we derive that
\begin{equation}
\left(\frac{1}{\nu} \frac{d^2\nu }{dr^2} + q_1\right)
\left(\frac{1}{\nu} \frac{d^2\nu }{dr^2} + q_2\right) -
\frac{t^{2}_{c}}{\gamma^4} \kappa = 0
\end{equation}
which can be written in the compact form as
\begin{equation}\label{polyn}
f^2 + (q_2 - q_1 ) f -  \frac{t^{2}_{c}}{\gamma^4} \kappa = 0,
\end{equation}
with the function $f$ being
\begin{equation}
f = \frac{1}{\nu} \frac{d^2\nu }{dr^2} + q_1.
\end{equation}
This function contains all the QD parameters including the energy
that we need to compute. One possible way to proceed is to write
$f$ in the form
\begin{equation}
f = \frac{1}{2} (q_1 - q_2) \pm \sqrt{
\left(\frac{q_1-q_2}{2}\right)^2 + \frac{t^{2}_{c}}{\gamma^4}
\kappa},
\end{equation}
which is a formal solution to Eq.~(\ref{polyn}). The last two
equations indicate that $\nu$ should satisfy the second-order
differential equation:
\begin{eqnarray}\label{approx1}
\frac{d^2\nu }{dr^2} + Q_{\pm} \nu = 0,
\end{eqnarray}
with the coefficient
\begin{equation}\label{qfull}
Q_{\pm}(r) = \frac{1}{2}(q_1+q_2) \\
\pm \sqrt{ \left(\frac{q_1-q_2}{2}\right)^2 +
\frac{t^{2}_{c}}{\gamma^4} \kappa}.
\end{equation}
In Sec.~\ref{compa} we use Eq.~(\ref{approx1}) to derive the
approximate energies of the QD. We usually need to consider only
$Q_{+}$, since $Q_{-}<0$ everywhere along $r$. But, whether we
have to account for $Q_{-}$ depends on the exact parameters
defining the QD as well as the energy range of interest.

For completeness we now follow a slightly different strategy to
derive an approximate QD equation similar to Eq.~(\ref{approx1}).
Starting with Eqs.~(\ref{w}) and (\ref{n}) and performing the
transformation
\begin{equation}\label{trans}
\left(\begin{array}{c}
 w \\
 \nu\\
\end{array}\right) = \frac{1}{2}\left(\begin{array}{cc}
   1  & 1 \\
   1  & -1  \\
\end{array}\right)
\left(\begin{array}{c}
  \chi_+ \\
  \chi_- \\
\end{array}\right),
\end{equation}
leads to the following coupled equations
\begin{subequations}
\begin{eqnarray}
\left(\frac{d^2}{dr^2} + \tilde{Q}_{-} \right)\chi_{+} = \frac{1}{2}(q_2 - q_1 ) \chi_{-}, \label{h-}\\
\left(\frac{d^2}{dr^2} + \tilde{Q}_{+} \right)\chi_{-} =
\frac{1}{2}(q_2 - q_1 ) \chi_{+}, \label{h+}
\end{eqnarray}
\end{subequations}
with
\begin{equation}\label{Qs}
\tilde{Q}_{\pm}(r) = \frac{1}{2}(q_2+q_1) \pm \frac{t_c}{\gamma^2}
[(E-V_1)(E-V_2)]^{1/2}.
\end{equation}
Equations~(\ref{h-}) and (\ref{h+}) are exact and suggest that
$\chi_{+}=w +\nu \approx 0$ when $q_2\approx q_1$ so that both
sides of Eq.~(\ref{h-}) simultaneously vanish. As a result, with
$\chi_{-} \approx 2\nu$ the final approximate equation for $\nu$
becomes a second order equation identical to Eq.~(\ref{approx1}),
but with $Q_{+}$ replaced by $\tilde{Q}_{+}$. Our numerical
calculations demonstrate that in general $Q_{+}$ gives more
accurate approximate results and for this reason we use $Q_{+}$
throughout this work. In Appendix~\ref{wkb} we further investigate
the approximate model and clarify in a rigorous way how the
condition $q_2 \approx q_1$ is essentially related to the
assumption that $\lambda$ is constant.

\begin{figure}[t]
\includegraphics[width=3.5cm, angle=270]{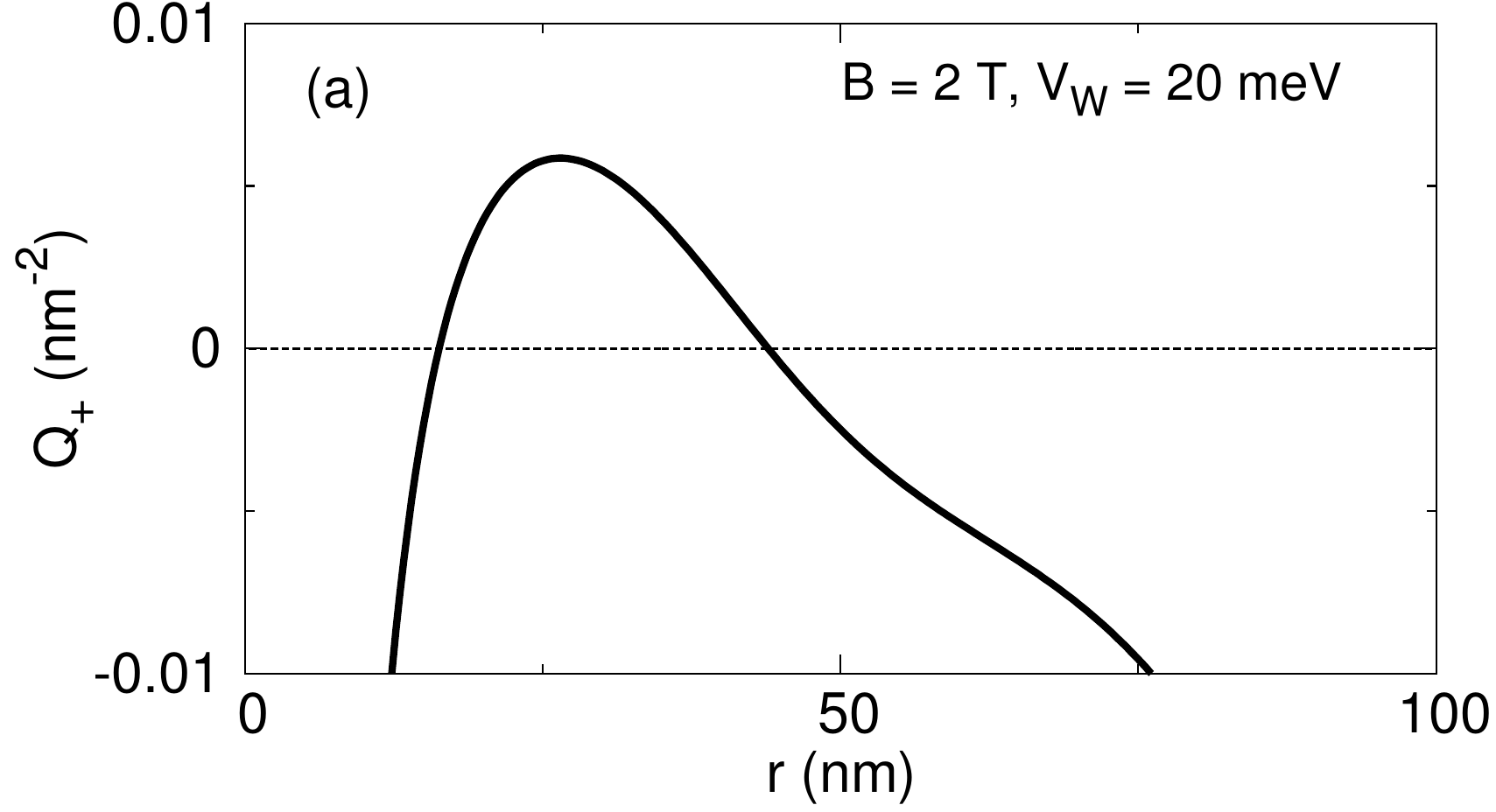}
\includegraphics[width=3.5cm, angle=270]{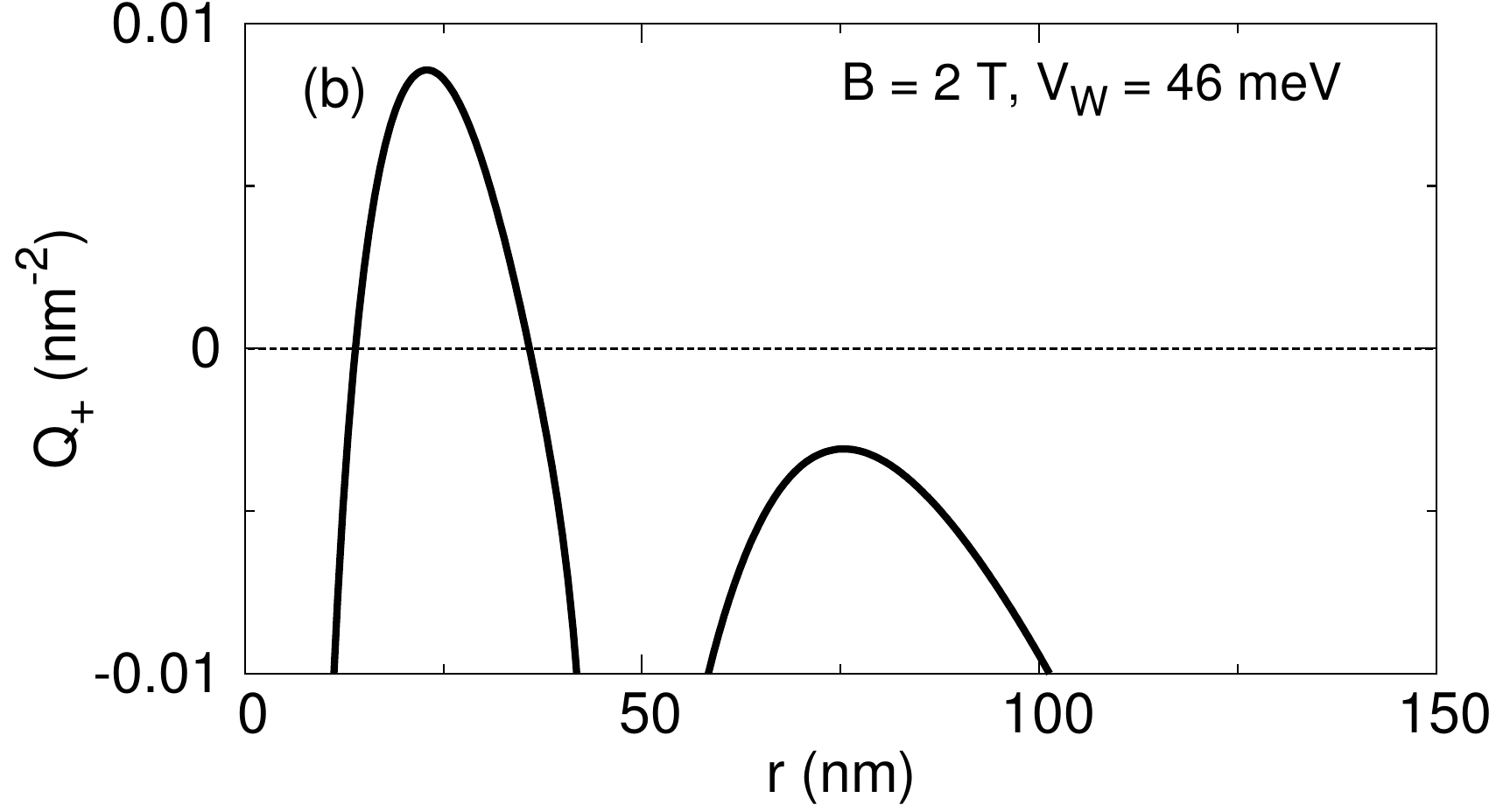}
\includegraphics[width=3.5cm, angle=270]{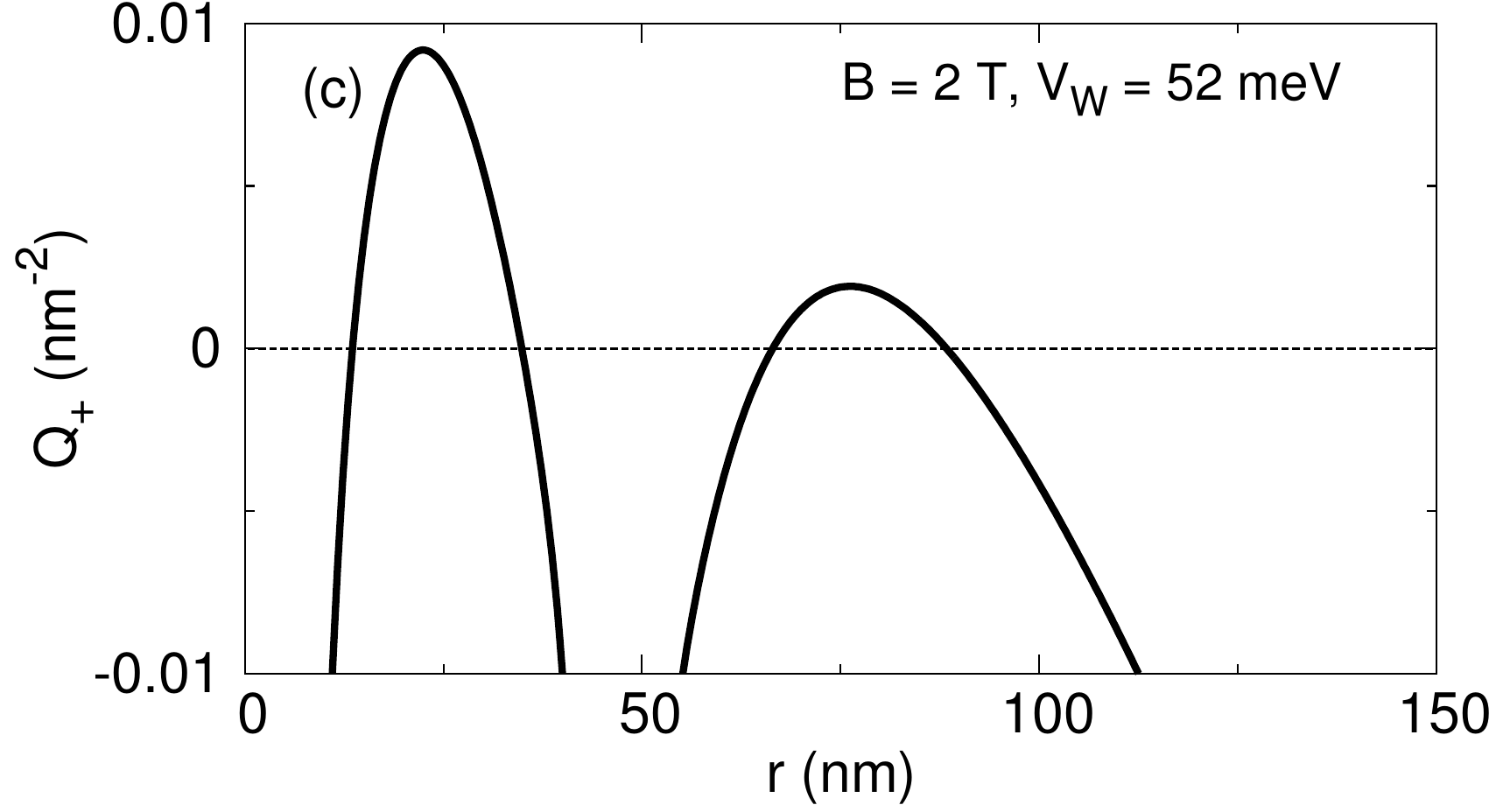}
\includegraphics[width=3.5cm, angle=270]{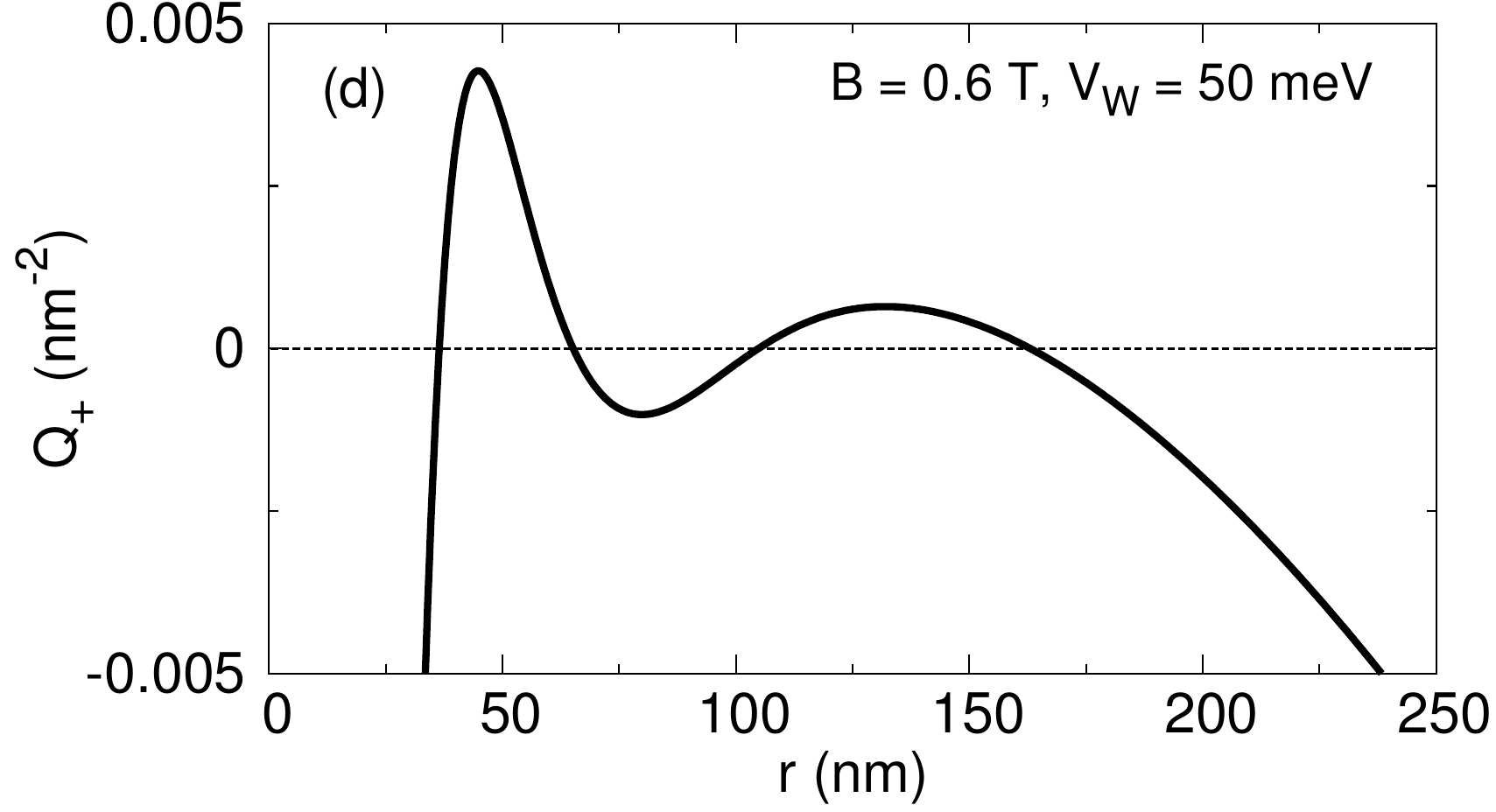}
\caption{The function $Q_{+}$, defined in Eq.~(\ref{qfull}), as a
function of radial coordinate at different magnetic fields and
potential wells for $s=2$, $V_{\rm b}=0$, and: (a) $m=2$,
$E\approx 1$ meV, (b) $m=2$, $E\approx - 20$ meV, (c) $m=2$,
$E\approx -25$ meV, and (d) $m=8$, $E\approx 15$
meV.}\label{testQ}
\end{figure}

We can numerically solve Eq.~(\ref{approx1}) to find the
approximate $E$ and $\nu$. However, using general arguments
pertinent to confined quantum states~\cite{schiff}, which have
been found to be applicable to monolayer
graphene~\cite{giavaras10, giavaras21}, we can make some further
approximations. Specifically, considering that $Q_{+}$ determines
a physically acceptable state $\nu$, we can define the region of
localisation from the condition $Q_{+}>0$. The simplest regime
occurs at $V_{\rm b}=0$ and when the singular point in $q_1$ and
$q_2$, satisfying $E-V_{\rm QD}=0$, can be safely ignored. We
focus on a QD parameter range where $Q_{+}$ has a single maximum
away from $r=0$ and retain terms up to $r^2$. We then expand $Q_+$
about $r_0$
\begin{equation}\label{Qminus}
Q_{+}(r) \approx Q_{+}(r_{0}) + \frac{1}{2} Q^{''}_{+}(r_0)
(r-r_{0})^{2},
\end{equation}
where $r_0$ is the position of the maximum and $Q^{''}_{+}(r_{0})$
is the second derivative of $Q_{+}(r)$ at $r_0$, with
$Q^{''}_{+}(r_{0})<0$. An important observation is that $r_0$ is
generally energy dependent, $r_0=r_0(E)$, and sensitive to the QD
parameters. Equations~(\ref{approx1}) and (\ref{Qminus}) indicate
that $\nu$ obeys a single-component Schr\"odinger equation similar
to that of a one-dimensional harmonic-oscillator located at $r_0$.
Thus, the energy $E$ of a physically acceptable confined state
$\nu$ should satisfy
\begin{eqnarray}\label{oscil}
\frac{Q_+(r_{0})}{ \sqrt{  \frac{1}{2} |Q^{''}_{+}(r_{0})| } } =
2n + 1, \quad n=0, 1 \ldots
\end{eqnarray}
where the integer $n$ is the number of nodes of the state $\nu$.
Equation~(\ref{oscil}) can be solved numerically, e.g., with
bisection, to find the energy $E$ for the corresponding value of
$n$. Compared with the four-component continuum model
Eq.~(\ref{oscil}) provides an approximate but much simpler way to
extract the QD energy levels and explore how these QD levels
emerge from the bulk Landau levels. The nodeless quantum state
$\nu$ has the exponential dependence
\begin{equation}
\nu(r) = e^{-(r-r_0)^2/2  \sqrt{|Q^{''}_{+}(r_{0})|/2}  },
\end{equation}
while ``excited'' states can be written in terms of Hermite
polynomials~\cite{schiff}. The number of nodes of a QD state in
monolayer graphene is sensitive to the position of the QD level
relative to the potential well~\cite{giavaras22}. It is
straightforward to show that this relativistic effect occurs also
in a bilayer graphene QD but is not captured by the approximate
Eq.~(\ref{oscil}). Consequently, within our approximate model a QD
state, when $V_{\rm QD}\ne 0$, has the same number of nodes as the
original Landau state. We return to this point in
Sec.~\ref{compa}.

\begin{figure}
\includegraphics[width=4.2cm, angle=270]{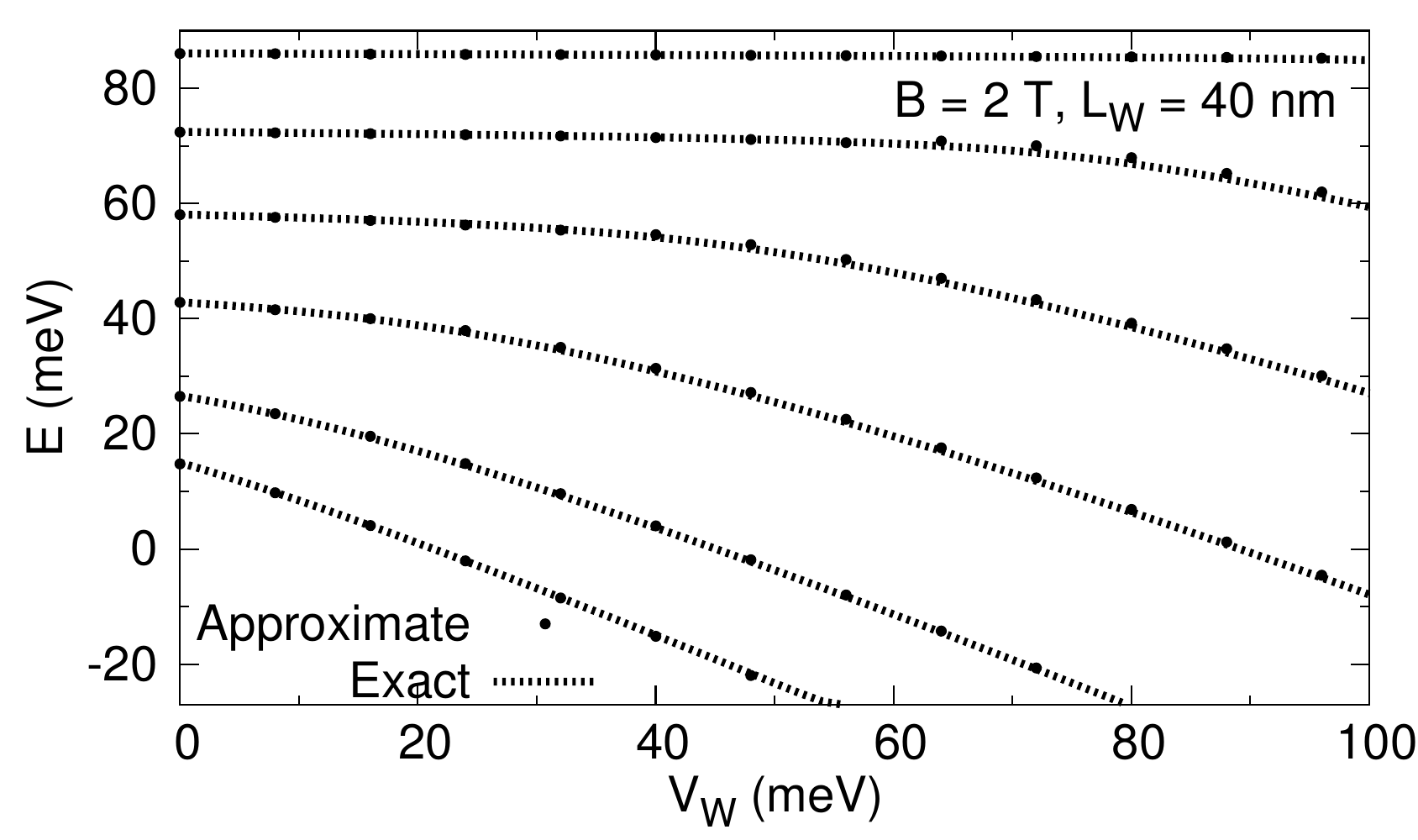}
\includegraphics[width=4.2cm, angle=270]{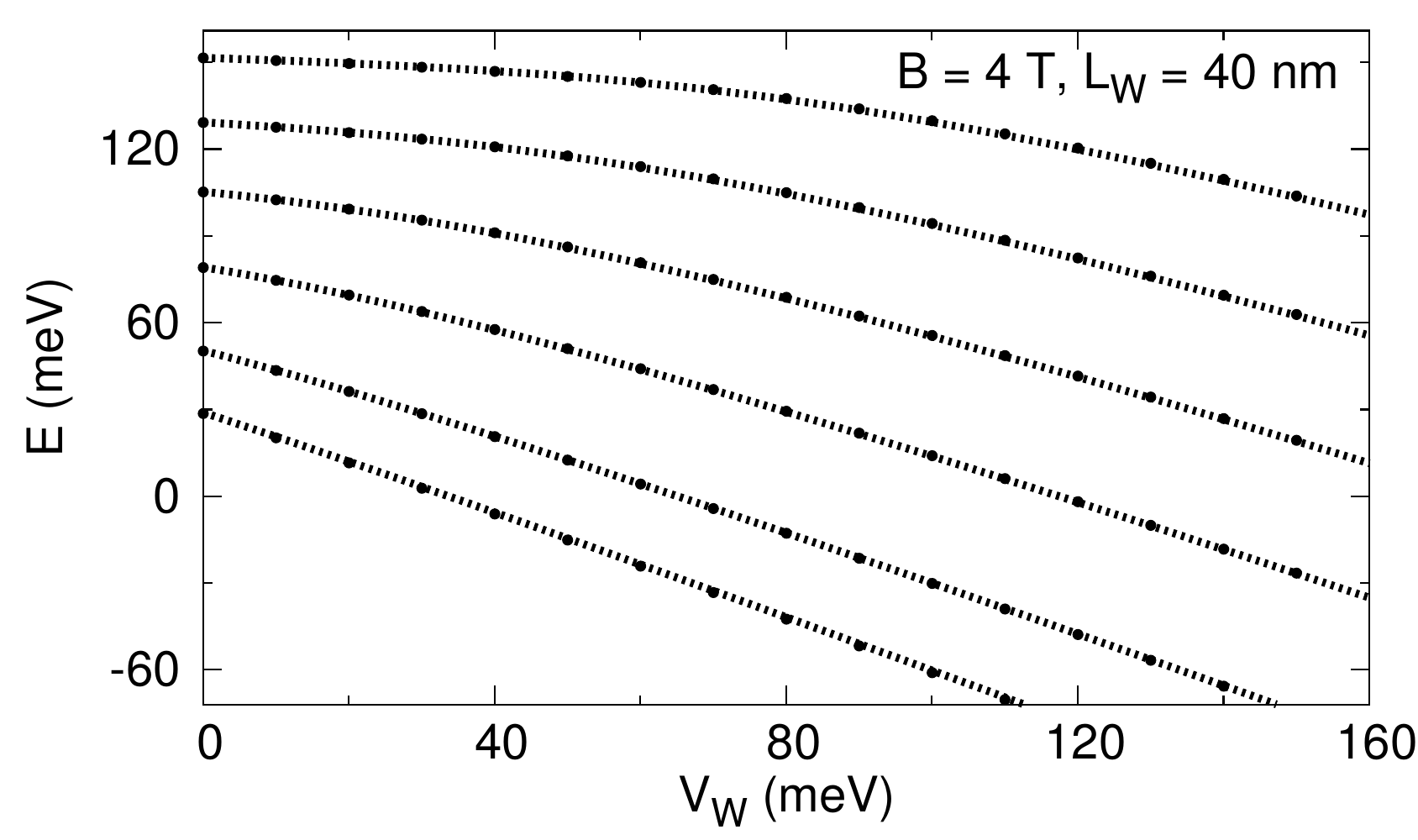}
\includegraphics[width=4.2cm, angle=270]{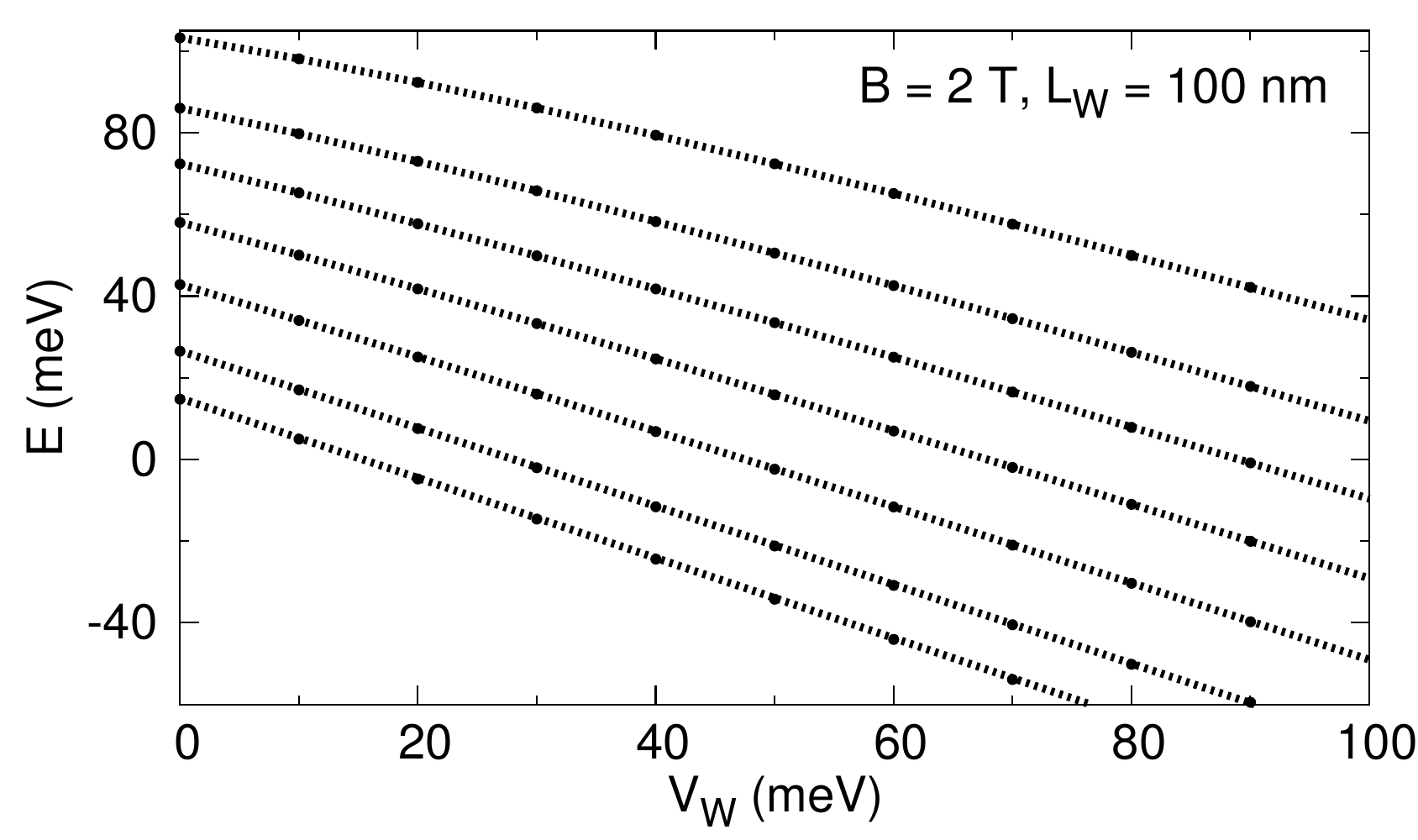}
\includegraphics[width=4.2cm, angle=270]{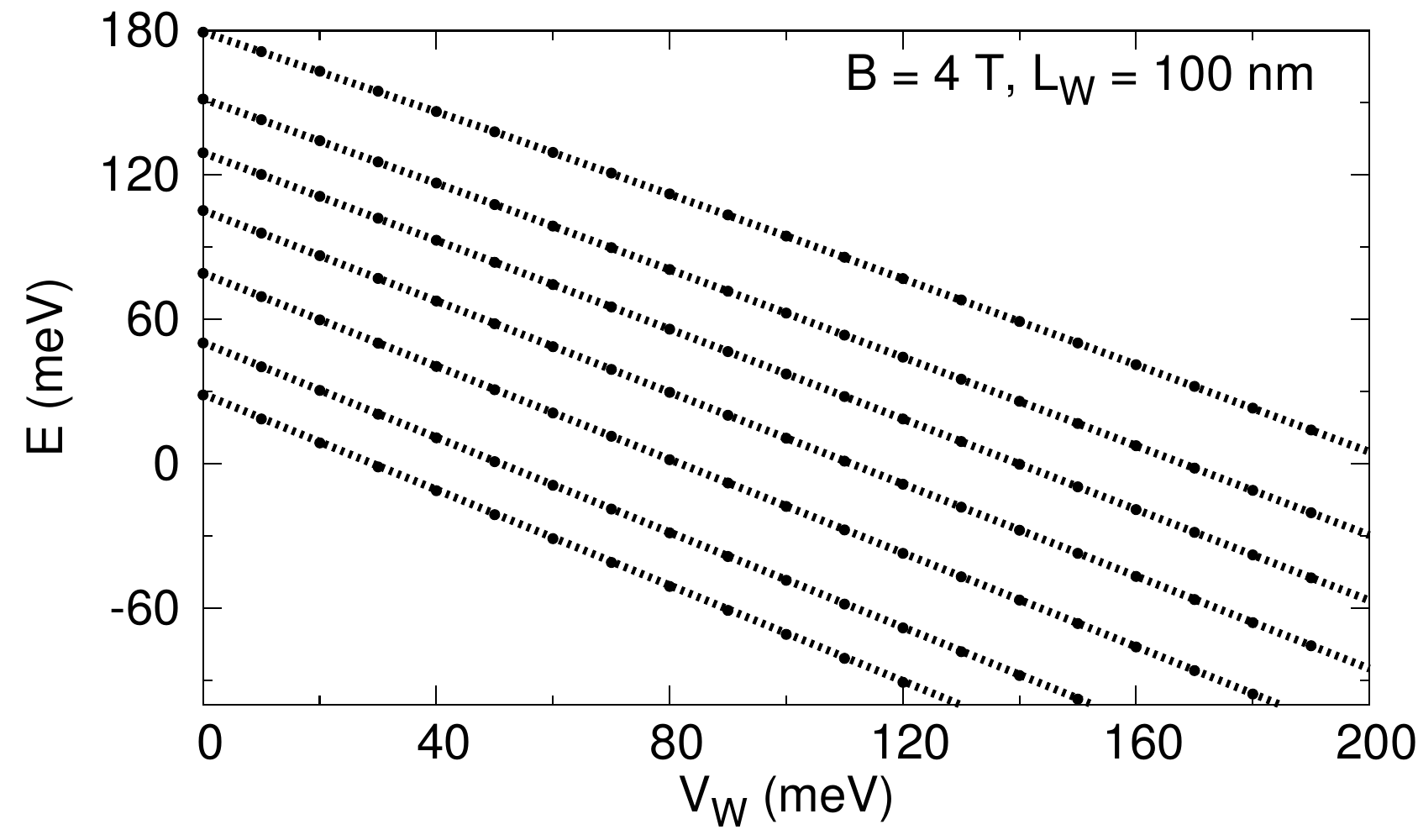}
\caption{Exact and approximate QD energy levels as a function of
QD potential well for $s=2$, and $V_{\rm b}=0$. From lower to
upper curve: $m=2$, 4, 7, 10, 13, 16, 20. Approximate QD levels
are derived from Eq.~(\ref{oscil}) for $n=0$.}\label{versusV}
\end{figure}

\section{Approximate quantum dot regimes}\label{cases}

In this section we describe the behavior of the function $Q_{+}$
[Eq.~(\ref{qfull})] that can arise in a QD, and specify the
approximate QD regimes. Because $Q_{+}$ can acquire a complicated
spatial dependence that sensitively depends on the QD parameters,
as well as the exact QD energy range, it is useful to identify
regimes with specific realistic examples.

In the simplest regime demonstrated in Fig.~\ref{testQ}(a) the
function $Q_{+}$ has a single positive maximum. Here, $Q_{+}$
arises for a QD energy which at $V_W=0$ corresponds to a positive
bulk Landau level, and in this ``single-maximum'' regime the
approximate energy can be derived from Eq.~(\ref{oscil}). In
Fig.~\ref{testQ}(a) we consider $V_{\rm b}=0$ and adjust the QD
parameters to give a QD energy greater than $V_{\rm asym}$, with
$V_{\rm asym} = V_{\rm QD}(r \rightarrow \infty)\approx 0$. It can
be readily shown that by gradually increasing $V_W$ the QD energy
decreases and crosses $V_{\rm asym}$, eventually resulting in a
well-defined negative maximum in $Q_{+}$ [see
Fig.~\ref{testQ}(b)]. A key observation is that the profile of the
function $Q_{+}$ continues to change at even larger values of
$V_W$ when the QD energy forms anticrossing points with energies
that originally correspond to negative Landau levels (hole
levels). This QD configuration, characterized by anticrossings for
$E<V_{\rm asym}$, has been demonstrated in monolayer
graphene~\cite{giavaras09, giavaras12} and to obtain the
corresponding approximate QD energies a different method is needed
from the one employed here.

Figure~\ref{testQ}(c) presents $Q_{+}$ for a typical case where
anticrossing points can be observed in the QD energy spectrum for
energies $E<V_{\rm asym}$ (Appendix~\ref{App}). Assuming $V_{\rm
b}=0$ the function $Q_{+}$ has now two positive maxima separated
by a central region where the singular point at $r\approx 44 $ nm
dominates. According to the above remarks the approximate method
developed in Sec.~\ref{first} becomes less accurate as $V_W$
increases and the QD levels start to form anticrossings. The
transition from the regime shown in Fig.~\ref{testQ}(a) to that in
Fig.~\ref{testQ}(c) can be observed, for example, by increasing
$V_W$. Between the two regimes Eq.~(\ref{oscil}) may still be
used, provided the inner maximum of $Q_{+}$ dominates. This
implies that the QD state has negligible amplitude around the
outer maximum, hence, this configuration is more likely to occur
for small values of $V_W$ (Sec.~\ref{compa}).

Figure~\ref{testQ}(d) shows another QD regime that arises when
$Q_{+}$ has again two positive maxima. The important, however,
difference from the example in Fig.~\ref{testQ}(c) is that now
there is no singular point, i.e., the QD energy is greater than
$V_{\rm asym}$. In the regime shown in Fig.~\ref{testQ}(d) the
approximate QD energy is obtained from the numerical solution of
Eq.~(\ref{approx1}), and as demonstrated in Sec.~\ref{compa} the
corresponding QD energy levels anticross for energies $E>V_{\rm
asym}$. Despite the simplicity of the approximate model the
anticrossing points can be accurately predicted by
Eq.~(\ref{approx1}). We emphasize that anticrossings are very
common in the energy spectra of graphene QDs~\cite{giavaras09,
giavaras12}, and in the present work we demonstrate how the
position and the size of these anticrossings can be directly
inferred from a simple inspection of the function $Q_{+}$.

Another approximate regime occurs for small $m$ values and when
$B$ as well as $L_W$ are large, so that the magnetic length $l_B =
( \hbar / e B)^{1/2}$ to be smaller than the effective width of
the potential well. In this regime the state is localised near the
center of the QD, $r \approx 0$, and to a good approximation the
QD energy $E$ decreases linearly with $V_{W}$. This can be
understood from the fact that in the region where the state is
localised the potential well induced terms in Eq.~(\ref{qfull})
containing derivatives of $V_{1}$ and $V_2$ are negligible.
Therefore, to a good approximation $V_{\rm QD}\approx - V_W$,
within the region where the state is localised. Focusing on a
positive Landau level $E_{\rm L}$, and taking $E=E_{\rm L}$ at
$V_W=0$ then by increasing $V_{W}$ we eventually expect the energy
$E$ to cross $V_{\rm asym} \approx 0 $. A simple way to determine
the linear dependence of $E$ versus $V_W$ is to use
Eq.~(\ref{oscil}) to find the approximate $V_W$ for which $E
\approx V_{\rm asym}$. By denoting this particular value of $V_W$
by $V^{0}_{W}$, the approximate energy as a function of $V_W$ can
be written as
\begin{equation}\label{linear}
E(V_W) \approx  E_{\rm L}  - E_{\rm L} \frac{V_W}{ V^{0}_{W} }.
\end{equation}
In this approximate treatment the maximum value of $V^{0}_{W}$ is
$E_{\rm L}$, therefore the linear drop cannot be greater than
$V_W$ as explicitly quantified in Sec.~\ref{compa} with various
numerical examples. This conclusion can be derived from a simple
inspection of Eq.~(\ref{oscil}). Specifically, if
Eq.~(\ref{oscil}) is satisfied for an energy $E$ corresponding to
a Landau level, $E=E_{\rm L}>0$ at $V_{W}=0$ (Landau regime), then
$E=0$ at $V_{W}=E_{\rm L}=V^{0}_{W}$ can also be a solution. At a
fixed angular momentum and magnetic field the limit $V^{0}_{W}
\approx E_{\rm L}$ can be achieved by increasing the effective
width $L_W$ of the QD.

\begin{figure}
\includegraphics[width=4.9cm, angle=270]{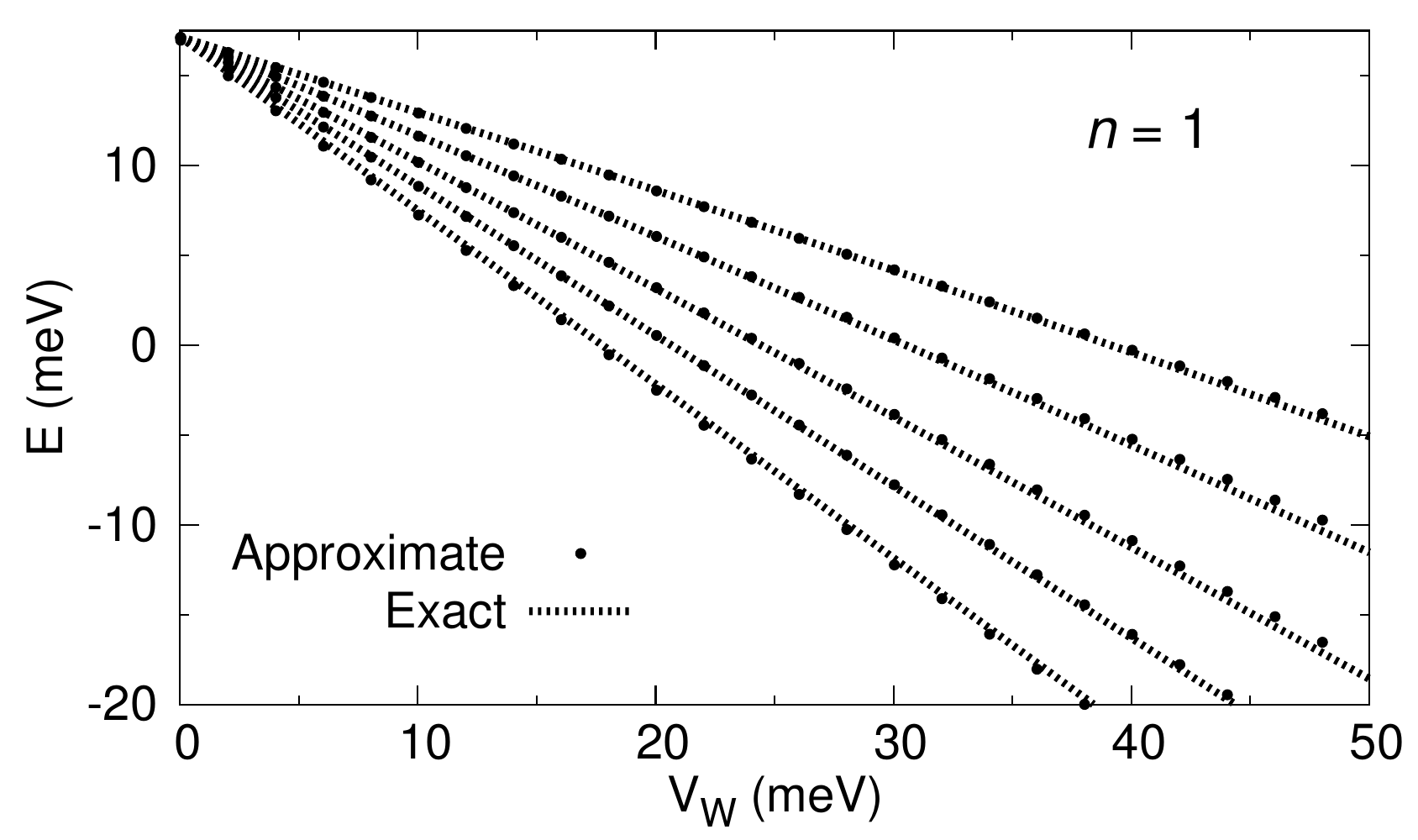}
\includegraphics[width=4.9cm, angle=270]{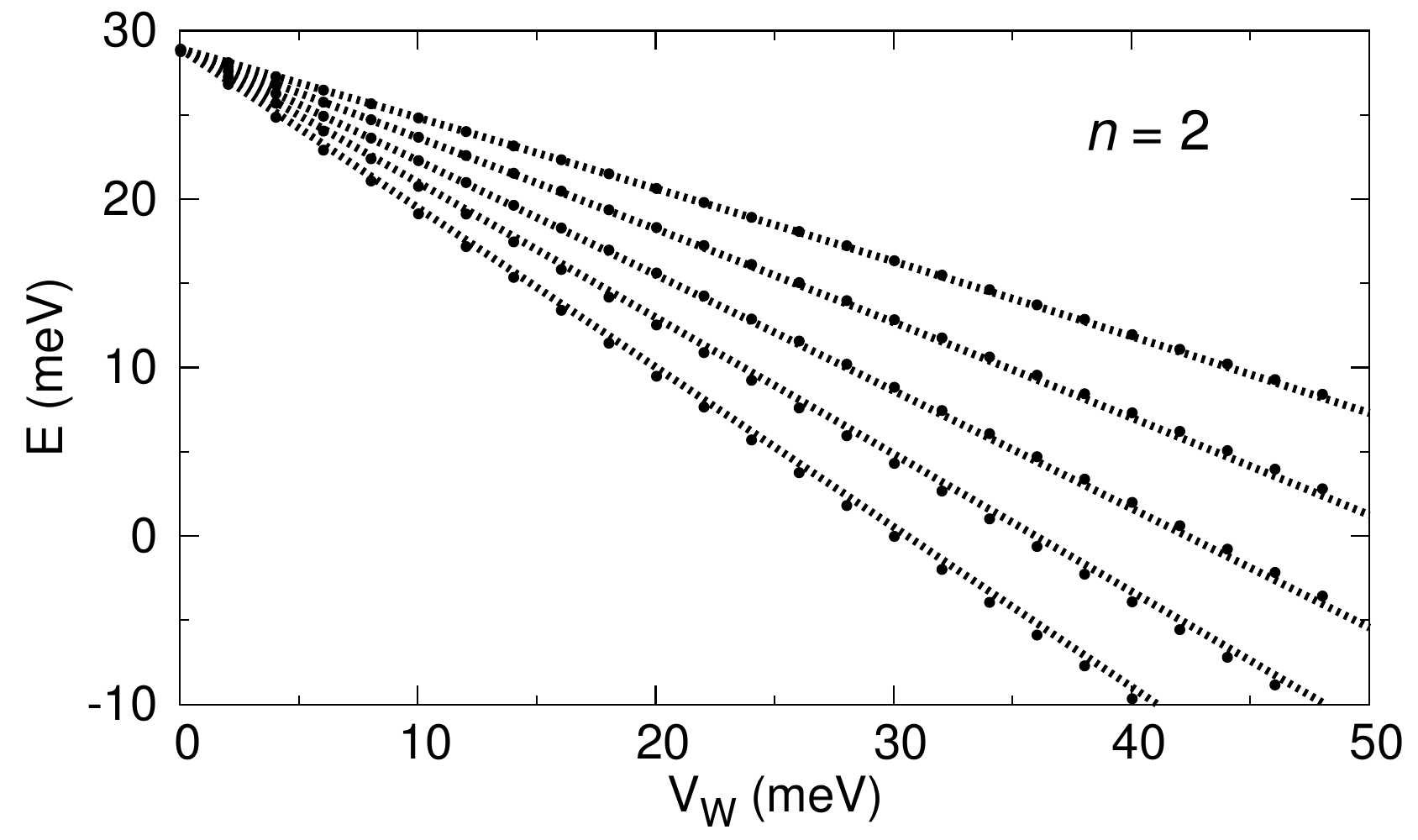}
\caption{Exact and approximate QD energy levels as a function of
QD potential well for $L_W=100$ nm, $B=4$ T, $s=2$, and $V_{\rm
b}=0$. From lower to upper curve: $m=-4$, $-16$, $-25$, $-35$,
$-45$. Approximate QD levels are derived from Eq.~(\ref{oscil})
for different $n$ values.}\label{negm}
\end{figure}

\begin{figure}
\includegraphics[width=5.5cm, angle=270]{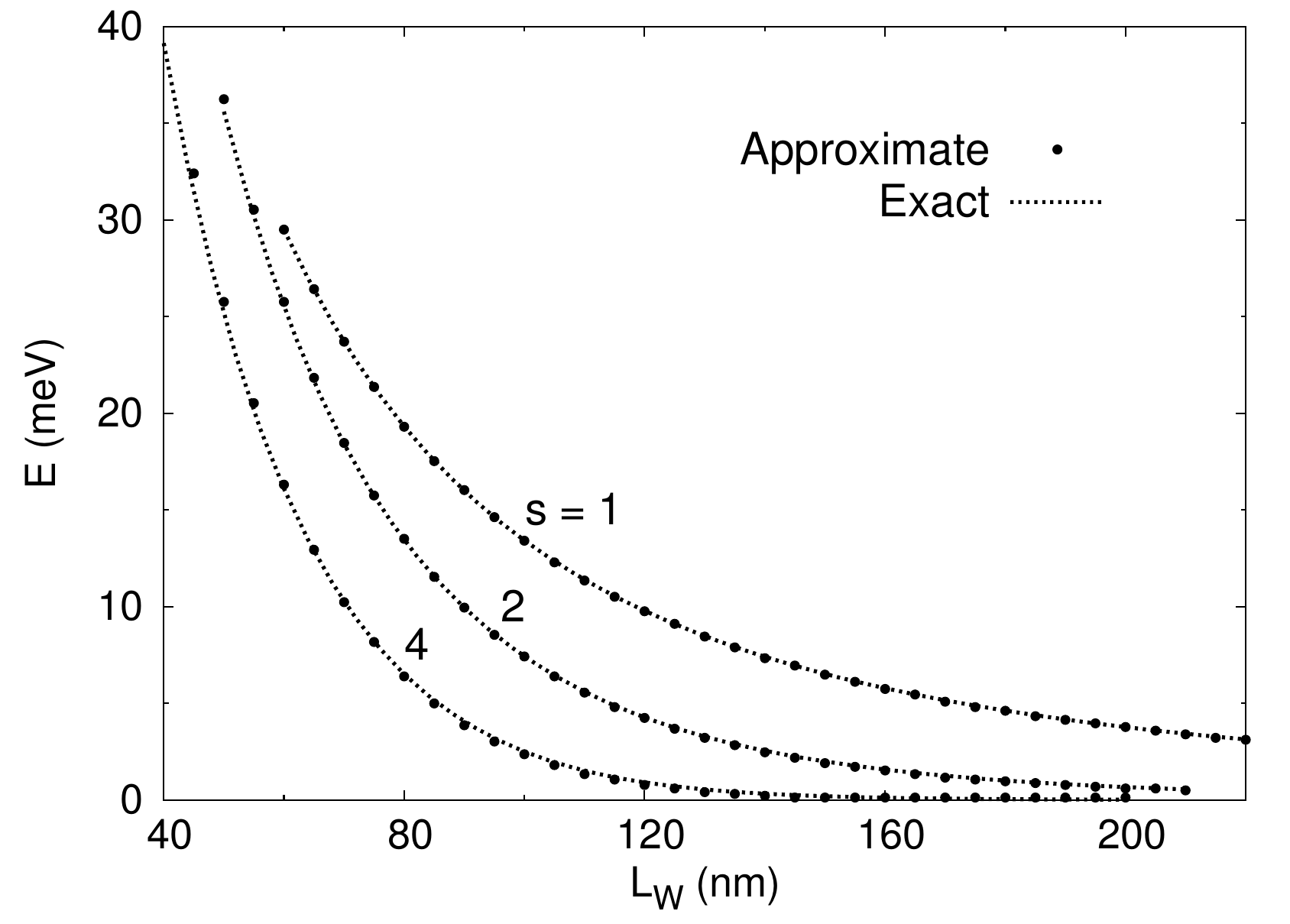}
\caption{Exact and approximate QD energy levels as a function of
QD potential width for $V_W\approx 48$ meV, $m=8$, $B=2$ T and
different superscripts $s$ in Eq.~(\ref{potential}). Approximate
QD levels are derived from Eq.~(\ref{oscil}) for $n=0$. For
clarity the curves for $s=1$ and $s=2$ are shifted by 20 nm and 10
nm respectively along the horizontal axis.}\label{zero}
\end{figure}

\begin{figure*}
\includegraphics[width=4.2cm, angle=270]{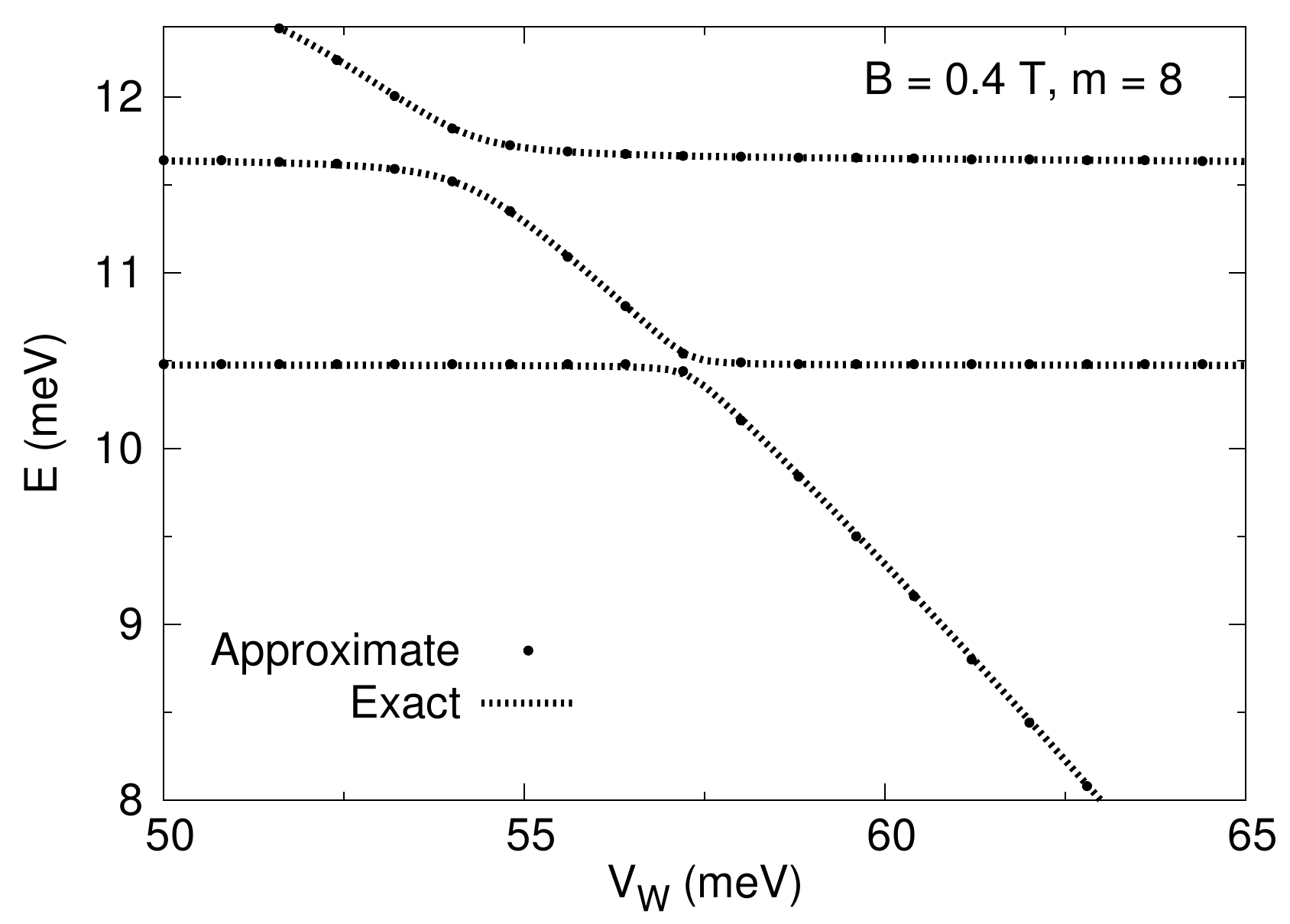}
\includegraphics[width=4.2cm, angle=270]{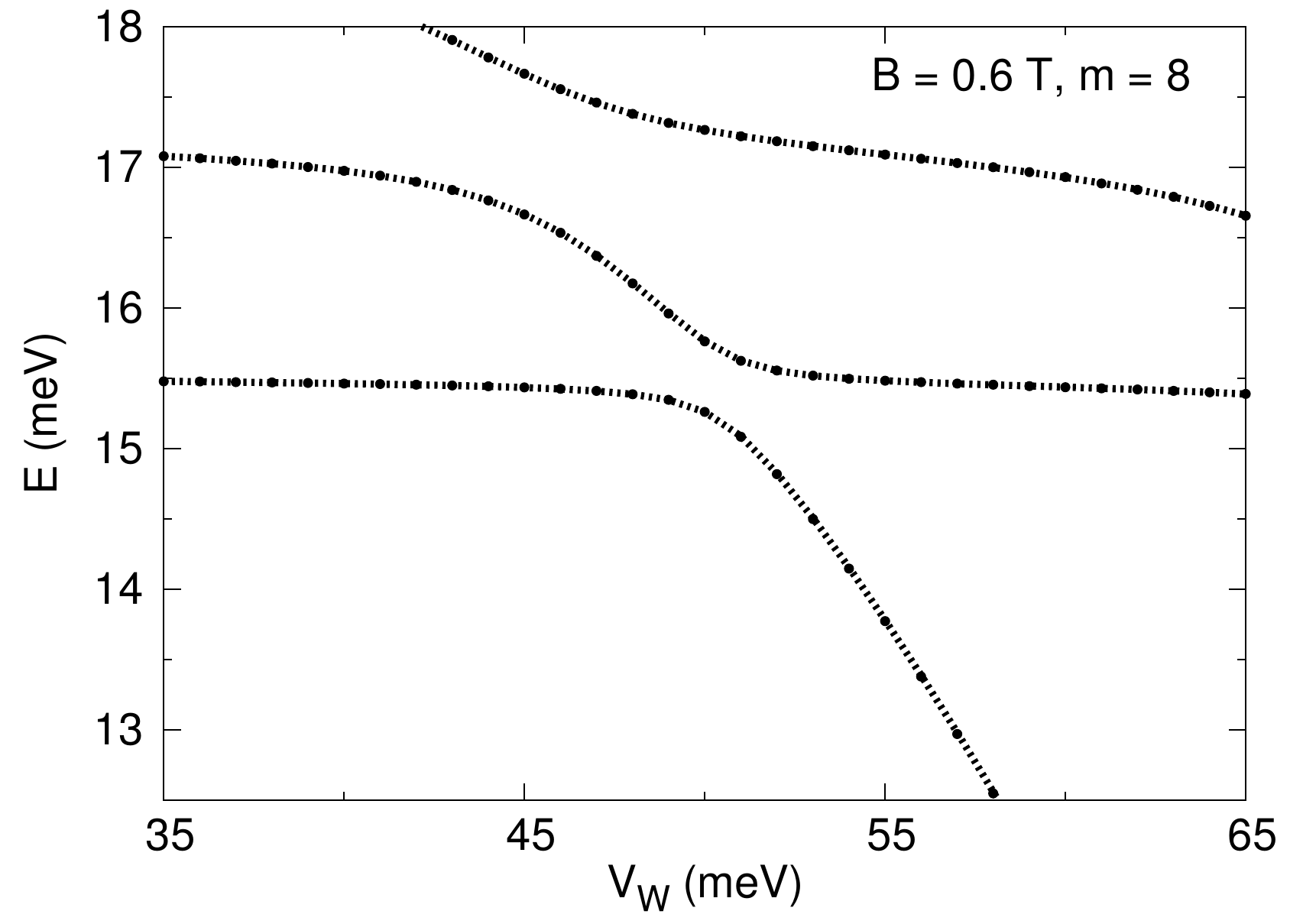}
\includegraphics[width=4.2cm, angle=270]{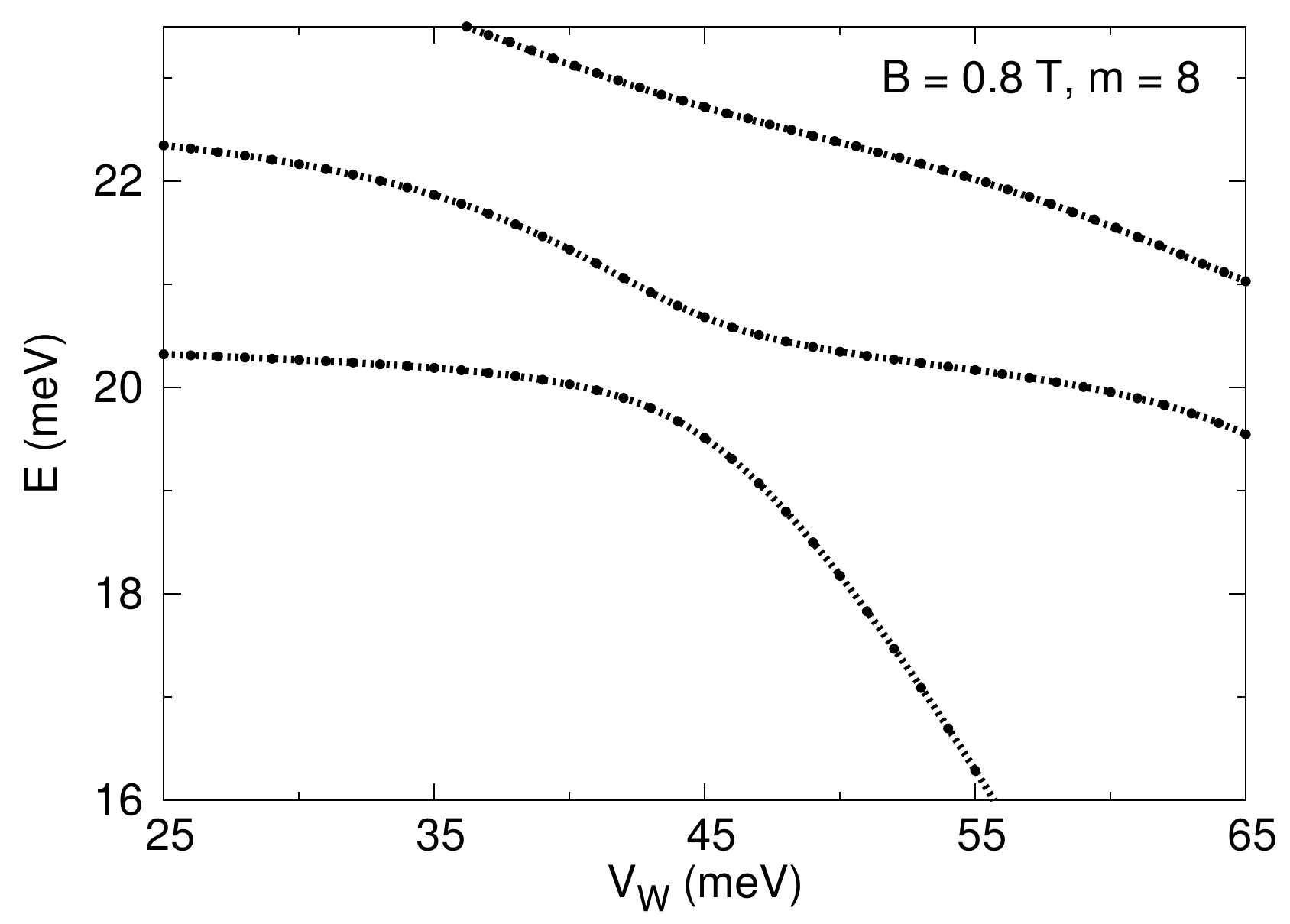}\\
\includegraphics[width=4.2cm, angle=270]{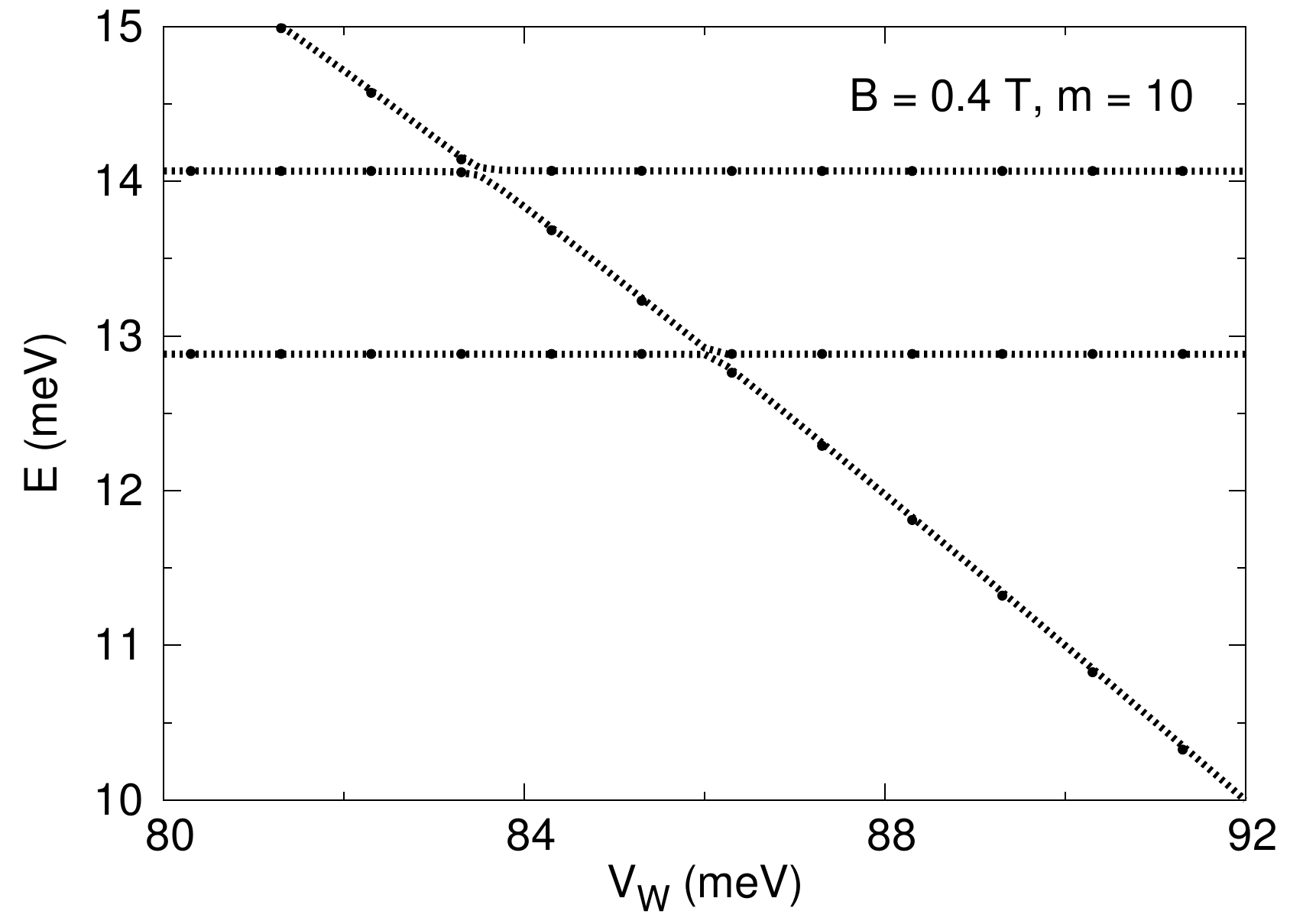}
\includegraphics[width=4.2cm, angle=270]{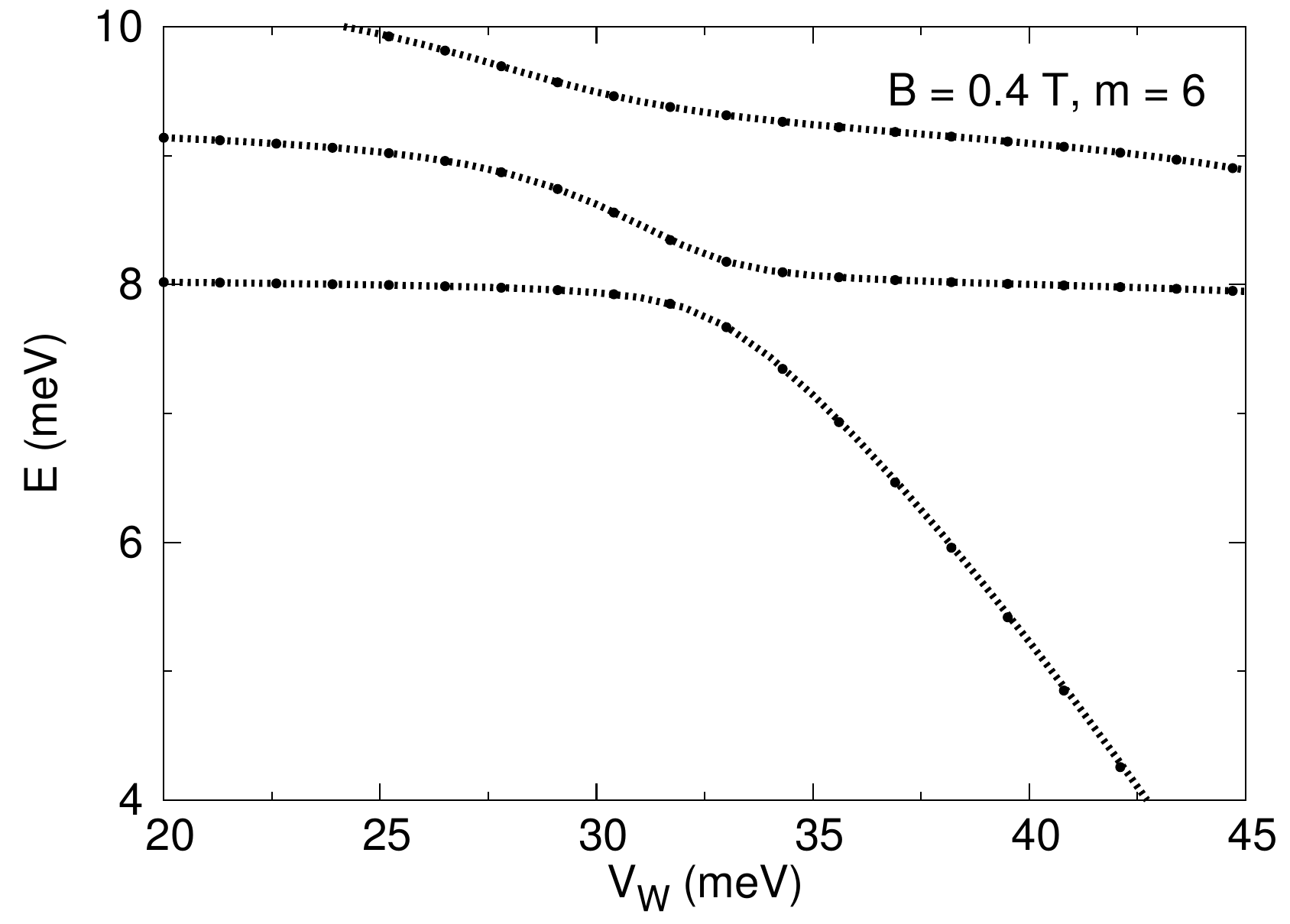}
\includegraphics[width=4.2cm, angle=270]{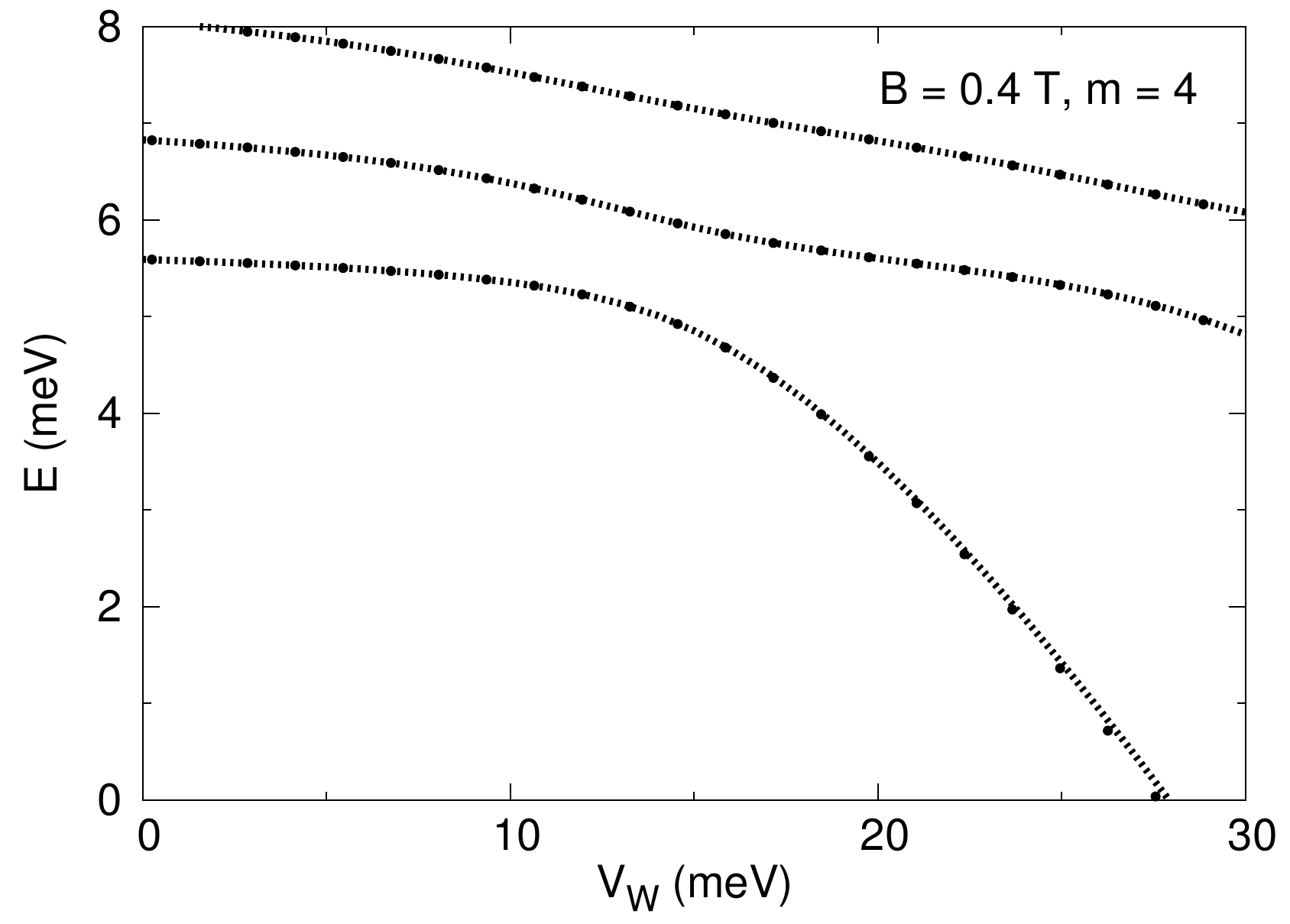}
\caption{Exact and approximate QD energy levels as a function of
QD potential well at different magnetic fields $B$ and angular
momenta $m$ for $s=2$ and $V_{\rm b}=0$. Approximate QD levels are
derived from the numerical solution of
Eq.~(\ref{approx1}).}\label{anti}
\end{figure*}

\section{Comparison between exact and approximate
energies}\label{compa}

In this section we quantify the approximations made in
Sec.~\ref{first} and compare the exact QD energy levels with the
approximate ones. The exact levels are derived from the numerical
solution of the four-component continuum QD model,
Eqs.~(\ref{dota})--(\ref{dotd}). The approximate levels are
derived from the single component approximate Eq.~(\ref{approx1}).
This is done either numerically, by solving Eq.~(\ref{approx1}),
or applying directly the harmonic oscillator formula,
Eq.~(\ref{oscil}).

For the numerical calculations of the exact QD levels we employ a
similar numerical approach to that in
Ref.~\onlinecite{giavaras09}. Some simple modifications are needed
to apply the appropriate boundary conditions that result in a
Hermitian eigenvalue problem~\cite{giavaras09}. In the present
work the derivatives in the radial Eqs.~(\ref{dota})--(\ref{dotd})
are approximated by a second-order finite-difference scheme with a
discretization length $\delta$. Compared with the first-order
scheme employed in Ref.~\onlinecite{giavaras09} the advantage is
that the numerical errors are of the order of $\delta^{2}$. The
disadvantage is that the resulting matrix is less sparse, however,
it can be easily treated using ARPACK subroutines incorporated
into a python program. For the numerical calculations we take
$\delta<0.2$ nm which ensures that the induced numerical errors
are small and do not affect the comparison between the exact and
the approximate QD levels.

The Landau regime ($V_{W}=0$) is the simplest regime to confirm
that Eq.~(\ref{oscil}) can be used to obtain approximate energies
for confined states. It can be easily verified by the analytical
form of the Landau states~\cite{peeters} that $w \approx \lambda
\nu$ for $t_c=400$ meV. Additionally, the function $Q_{+}$ defined
in Eq.~(\ref{qfull}) for $m\ne0$ has a single positive maximum at
$r_0$, and interestingly, $r_0$ is now energy independent. This
remark allows us to drastically simplify Eq.~(\ref{oscil}) and
perform analytical calculations to derive the Landau levels.
However, since the Landau levels have been analytically explored
in detail in an earlier work~\cite{peeters} we do not pursue here
a further investigation, and focus on the QD regime corresponding
to $V_W\ne0$.

Figure~\ref{versusV} shows some QD energy levels as a function of
the QD potential well $V_W$. The approximate levels have the
correct dependence on $V_W$ and agree very well with the exact
levels in a broad range of $V_W$ and different angular momenta
$m$. At $V_W=0$ the approximate method reproduces the bulk Landau
levels very accurately. When $V_W\ne0$, then in general larger $m$
levels are affected less by the potential well, and for a fixed
$m$ the role of $V_W$ is more significant at larger magnetic
fields~\cite{giavaras12, giavaras21}. Furthermore, in agreement
with the remarks in Sec.~\ref{first}, Fig.~\ref{versusV}
demonstrates that at larger magnetic fields the QD levels exhibit
an approximate linear dependence on $V_W$, which is more
pronounced for smaller $m$ values.

In Fig.~\ref{versusV}, focusing on a given energy and gradually
increasing $V_W$ shifts this energy within the potential well,
$E<V_{\rm asym} \approx 0$. According to the exact QD model,
Eqs.~(\ref{dota})--(\ref{dotd}), when such a shift takes place
extra nodes appear in the QD states as happens in monolayer
graphene~\cite{giavaras22}. The key observation is that when $V_W$
is not too large a node appears within a spatial region where the
QD state has low amplitude~\cite{giavaras22}. Therefore, the
formation of the extra node is not expected to alter significantly
the physics. For this reason the approximate harmonic oscillator
formula, Eq.~(\ref{oscil}), is still applicable and predicts the
corresponding energy very accurately.

In Fig.~\ref{versusV} we only plot QD levels which emerge from
positive bulk Landau levels, however, the approximate method can
also be successfully applied to QD levels that originate from
negative Landau levels (Appendix~\ref{App}). Similar to monolayer
graphene~\cite{giavaras12}, by increasing $V_W$ some QD energy
levels start to anticross for energies $E<V_{\rm asym}$. The
resulting QD levels arise from the hybridization between electron
and hole Landau levels, and in this regime the function $Q_{+}$
has the characteristic form shown in Fig.~\ref{testQ}(c). As
discussed in Sec.~\ref{first} this regime is beyond the scope of
the present work and cannot be captured by Eq.~(\ref{oscil}).

Figure~\ref{negm} presents some QD energy levels as a function of
the QD potential well for negative $m$ values. The agreement
between the approximate energy levels and the exact ones is again
very good especially for small values of $V_W$. Quantum dot levels
can also emerge from the zero-energy Landau level, however, for
these levels the assumption, $1/ \lambda = \nu/w$,
[Eq.~(\ref{lambda})] is not satisfied and Eq.~(\ref{oscil}) cannot
be applied. In this QD regime the singular point is expected to be
important and a different strategy is needed to obtain the
approximate QD energies. However, we can still obtain some
physical insight without significantly deviating from the analysis
in Sec.~\ref{first}. We can loosely assume that $\phi_A \gg
\phi_{B'}$ and then set $\chi_{\pm} \approx w$ in Eqs.~(\ref{h-})
and (\ref{h+}) to derive a single differential equation for $w$.
The resulting equation is similar to Eq.~(\ref{approx1}) with
$Q_{\pm}(r) = q_1(r) \pm [E-V_{\rm QD}(r)]t_c/\gamma^2$ when
$V_{\rm b}=0$. This approximate equation is particularly good for
small values of $m$ and $V_W$. As an example, at $V_W=31$ meV the
relative error is about $0.5\%$ for $m=-4$ but this quickly
increases to $5\%$ for $m=-16$.

The results in Fig.~\ref{versusV} and Fig.~\ref{negm} are for
$s=2$ and we have confirmed that good agreement between exact and
approximate energies is found for different values of $s$
(Appendix~\ref{App}). We briefly explore the role of $s$ in
Fig.~\ref{zero}, where we plot the QD levels as a function of the
potential well width $L_W$. Here, the potential well is fixed and
equal to $V_W \approx E_{\rm L}$, so for this reason the QD energy
tends to $E \approx 0$ with $L_W$. This behaviour is robust and
agrees with the approximate Eq.~(\ref{linear}). This confirms the
fact that the decrease of a QD level from the bulk Landau level
cannot be greater than $V_W$, and according to Fig.~\ref{zero} for
larger $s$ values the QD energy saturates at smaller potential
widths.

\begin{figure}
\includegraphics[width=4.5cm, angle=270]{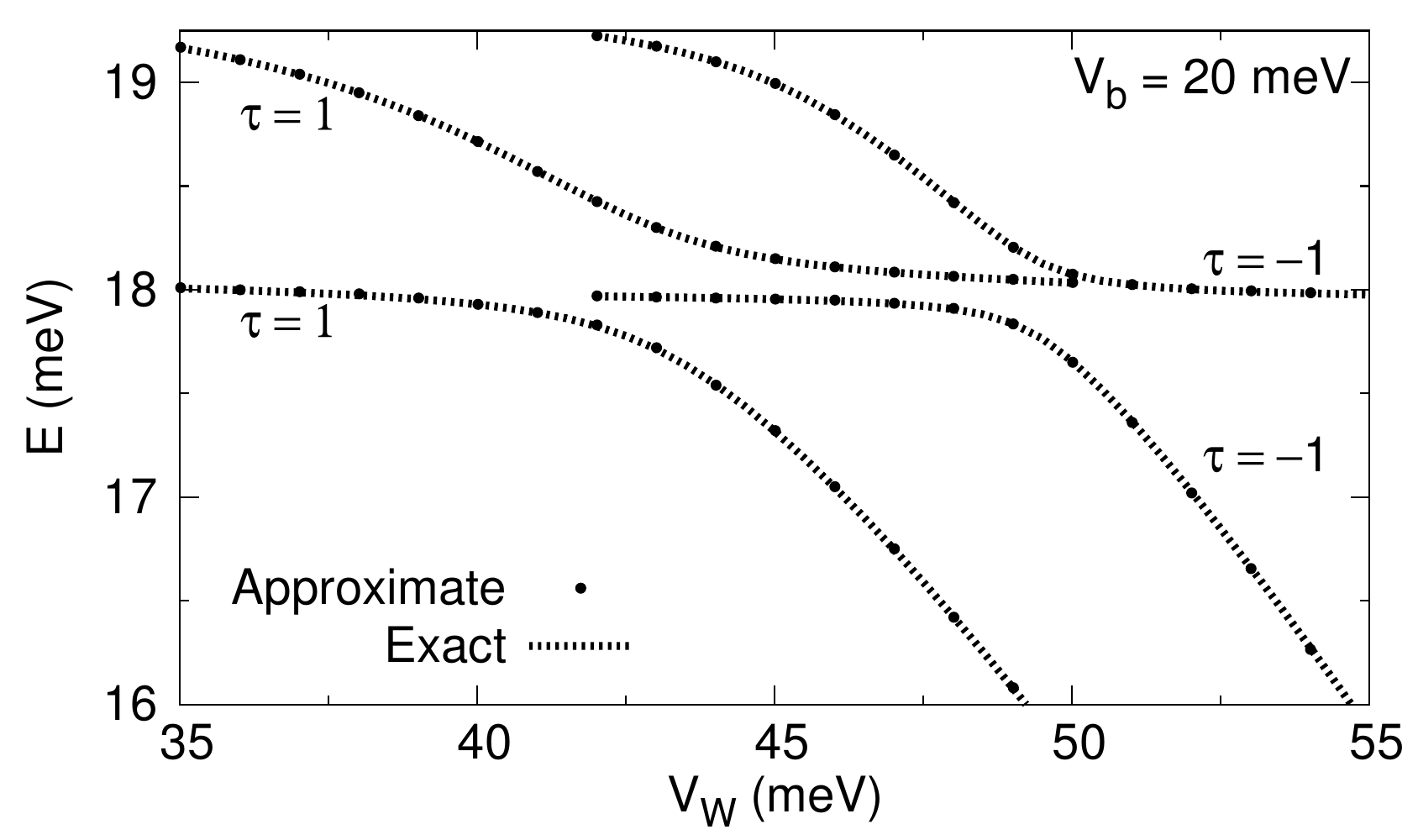}
\includegraphics[width=4.5cm, angle=270]{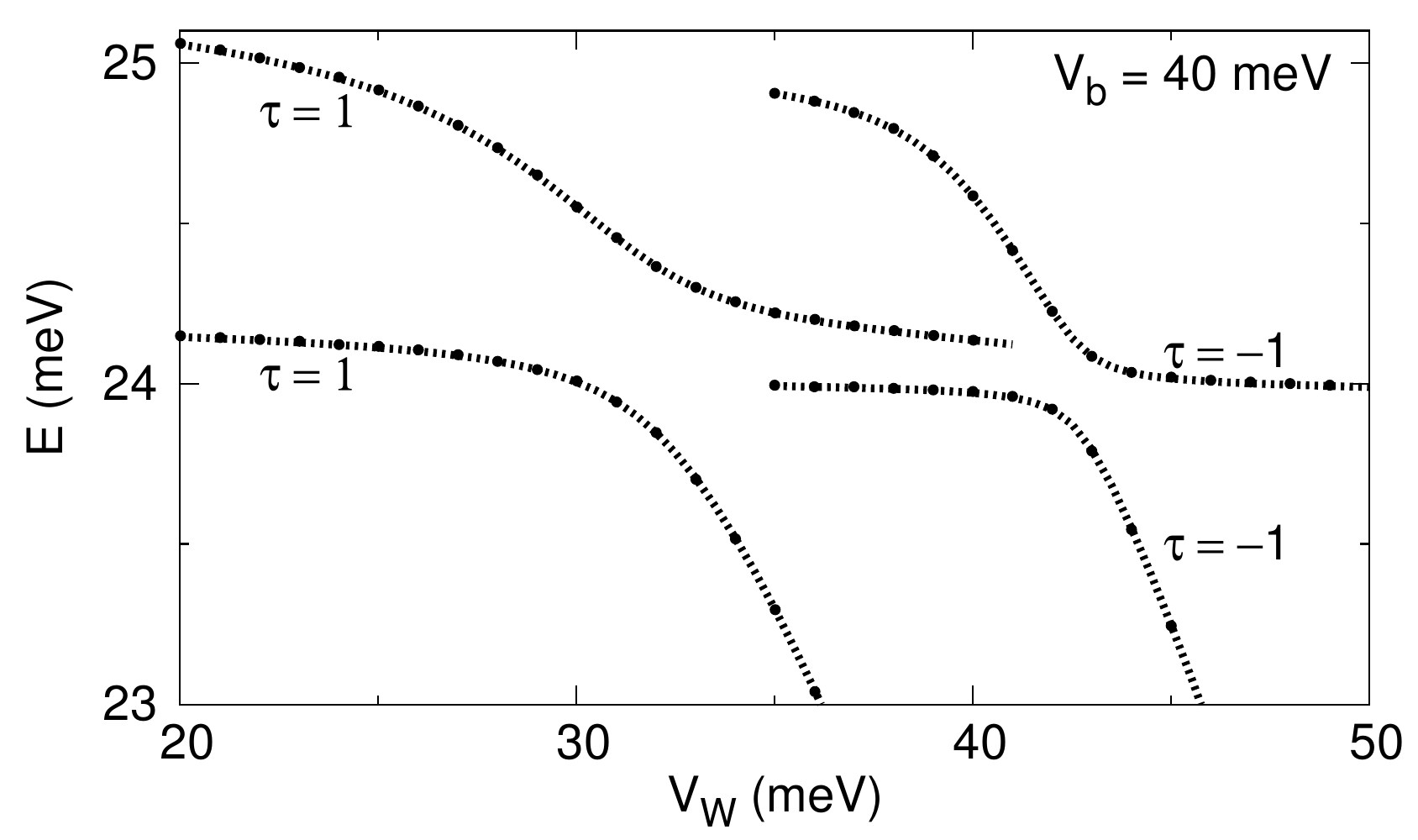}
\caption{Exact and approximate QD energy levels as a function of
QD potential well in biased BLG for $\tau=\pm 1$, $B=0.6$ T,
$m=8$, $L_W=40$ nm, and $s=2$. Approximate QD levels are derived
from the numerical solution of Eq.~(\ref{approx1}).}\label{bias}
\end{figure}

We now proceed to demonstrate another interesting regime which can
be approximately studied and seems to be mostly unexplored by
earlier exact numerical studies of QDs in BLG. This regime can be
more easily identified at low magnetic fields and potential
widths. The key requirement is the potential well region to be
different from the region where the original Landau state is
localised. A narrow potential well tends to confine the states
near the centre, $r\approx0$, whereas a small magnetic field away
from the centre. Tuning $V_W$ can give rise to anticrossing points
for energies $E>V_{\rm asym}$, and the function $Q_{+}$ has two
positive maxima as illustrated in Fig.~\ref{testQ}(d). To obtain
the approximate QD levels in this regime we perform a numerical
solution of Eq.~(\ref{approx1}) and present various examples in
Fig.~\ref{anti}. We focus on the few lowest anticrossing points
and choose the magnetic field to be relatively low, $B < 1$ T, so
that the Landau states to be localized away from the centre. The
results in Fig.~\ref{anti} indicate that both $B$ and $m$ should
have suitable values in order to resolve the anticrossings.

In Fig.~\ref{anti} the energy level that arises from the potential
well shifts downwards with $V_W$, in contrast the Landau level
depends weakly on $V_W$. As the $B$ field decreases the Landau
state shifts away from the centre, therefore, its overlap with the
state localised in the potential well weakens. Consequently, the
anticrossing gap starts to close as seen in Fig.~\ref{anti} at
$B=0.4$ T and $m=8$. The opposite situation occurs for smaller $m$
values for which the Landau state strongly overlaps with the
potential well state leading to a gap opening; see for example
Fig.~\ref{anti} at $B=0.4$ T and $m=4$. The fact that the
anticrossing is formed for $E>V_{\rm asym}$, thus $E-V_{1,2}\ne
0$, suggests that to a good approximation the QD energy levels can
be obtained from a simplified function $Q_{+}$. We have confirmed
that by neglecting all the terms containing derivatives of
$V_{1,2}$ in $Q_{+}$ [Eq.~(\ref{qfull})] reproduces the exact
energy characteristics.

Finally, in Fig.~\ref{bias} we illustrate the formation of an
anticrossing point in biased BLG for the two valleys. As in the
zero-bias case, the approximate QD levels agree very well with the
exact levels and we have confirmed that similar agreement is
achieved at different magnetic fields and QD parameters beyond the
anticrossing point regime. In biased BLG the function $Q_+$
defined in Eq.~(\ref{qfull}) can acquire an imaginary part since
the square-root can be negative when $\kappa<0$. Provided that
this imaginary part arises far away from the region where $Q_{+}$
is solely real, with $Q_{+}>0$, the approximate energies can be
obtained in the same way as for $V_{\rm b}=0$. The present work
focuses on this regime only, hence, we can numerically solve the
approximate Eq.~(\ref{approx1}) or use directly Eq.~(\ref{oscil})
to extract the approximate energies.

The accuracy of the harmonic oscillator formula,
Eq.~(\ref{oscil}), depends on how well the parabolic term can
approximate $Q_{+}$ in Eq.~(\ref{Qminus}). This, however, cannot
be predicted in advance (Appendix~\ref{App}) due to the many terms
involved in the approximate equations, as well as the broad range
of the various physical parameters. Our numerical investigation
suggests that the approximate results are usually less accurate
for negative angular momenta and, in general, for smaller absolute
values of angular momenta. For small energies, especially near
zero, the relative errors can be large but the correct
characteristics can still be predicted. In contrast, the relative
errors tend to be smaller at higher magnetic fields and larger QD
energies as well as for larger values of $L_W$. Additionally, the
accuracy of the approximate QD levels derived from
Eq.~(\ref{oscil}) can be improved by numerically solving the
approximate Eq.~(\ref{approx1}).

\section{Conclusion and discussion}\label{conclu}

We considered a QD in bilayer graphene defined by a realistic
continuous potential well in a constant magnetic field. We started
with the four-component continuum model and demonstrated some
regimes where the QD energy levels can be obtained from simplified
approximate equations. The basic advantage is that the approximate
equations involve a single-component wavefunction and thus provide
pedagogical insight into the physics of QDs. The approximate
equations are applicable to a general form of the QD potential
with either a soft-wall or hard-wall quantum well. The equations
clarify in a transparent way how the quantum dot confinement
changes by increasing the strength of the quantum well and the
magnetic field, as well as how the QD levels emerge from the bulk
Landau levels.

We proved the efficacy of the approximate method by comparing the
approximate energies with the exact energies derived from the
four-component continuum model in BLG. We found that the
approximate energy levels as a function of the magnetic field and
QD potential well exhibit the correct general features, such as
for example, anticrossing points as well as a linear dependence on
the dot potential, and agree very well with the exact energy
levels in a broad range of QD parameters. We demonstrated various
realistic regimes where the relative error can be vanishingly
small.

Semiclassical and quantum super-symmetric methods might be
interesting alternative frameworks~\cite{nayfeh} to approximately
explore QD physics in bilayer graphene. Equation~(\ref{approx1})
can be used as the starting point of a WKB method~\cite{schiff,
nayfeh} with $\sqrt{Q_{+}}$ defining the local wavelength. The
approximate QD energies can be obtained from a standard
Bohr-Sommerfeld quantization condition involving the classical
turning points, $Q_{+}=0$. Each method is expected to have some
advantages in terms of numerical effort, accuracy, and pedagogical
insight. The harmonic oscillator formula derived in the present
work is easily implementable and accurate enough. An interesting
case which seems to remain unexplored corresponds to the quantum
dot regime where a singular point separates two classically
allowed regions [Fig.~\ref{testQ}(c)]. This configuration that
couples electron and hole levels cannot be captured by our
approximate model. A more powerful QD model should treat equally
well the different regimes illustrated in Fig.~\ref{testQ}.

\begin{acknowledgments}
I wish to thank the anonymous referees for some useful comments.
\end{acknowledgments}

\setcounter{secnumdepth}{0} 

\setcounter{secnumdepth}{1}

\begin{figure}
\includegraphics[width=4.3cm, angle=270]{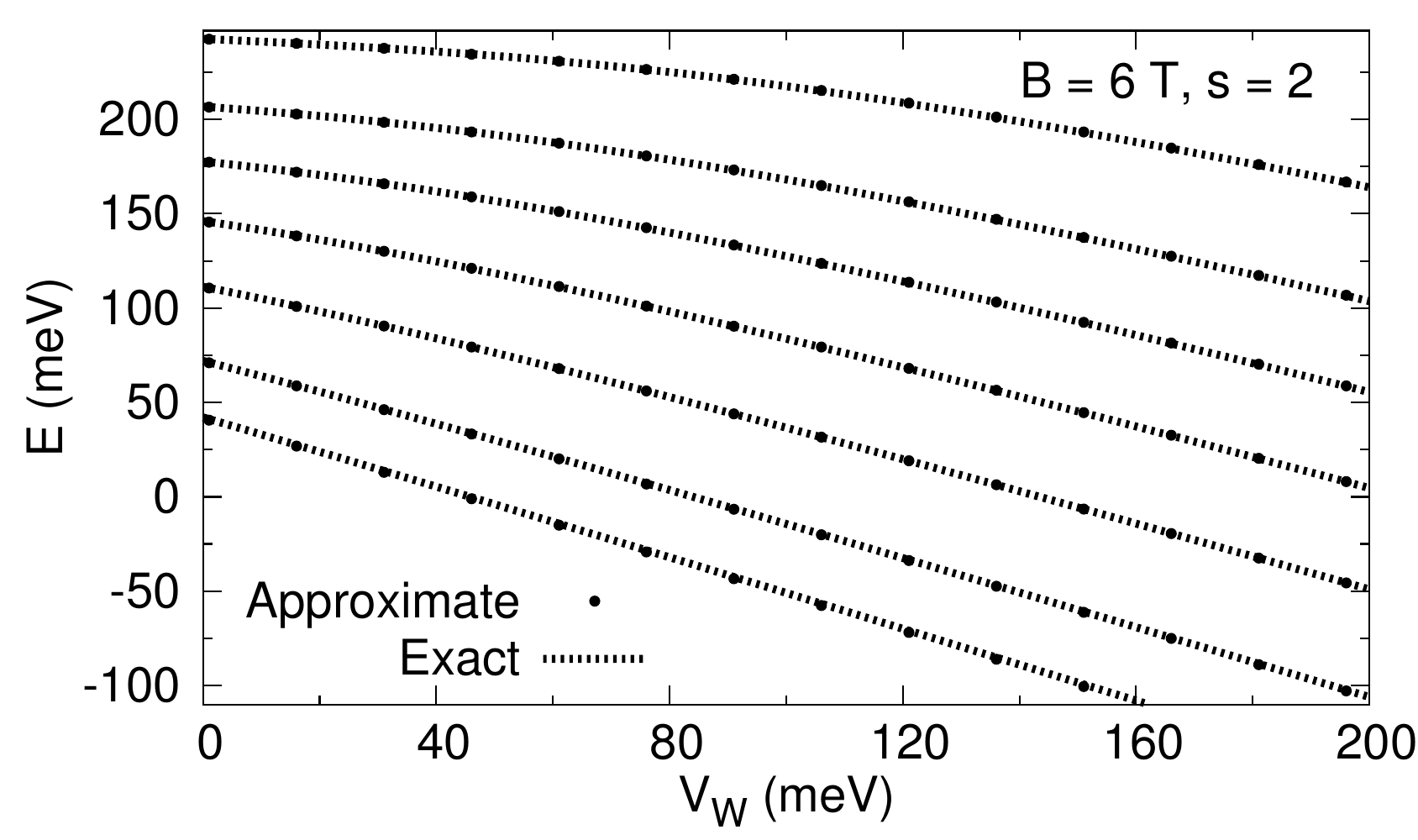}
\includegraphics[width=4.3cm, angle=270]{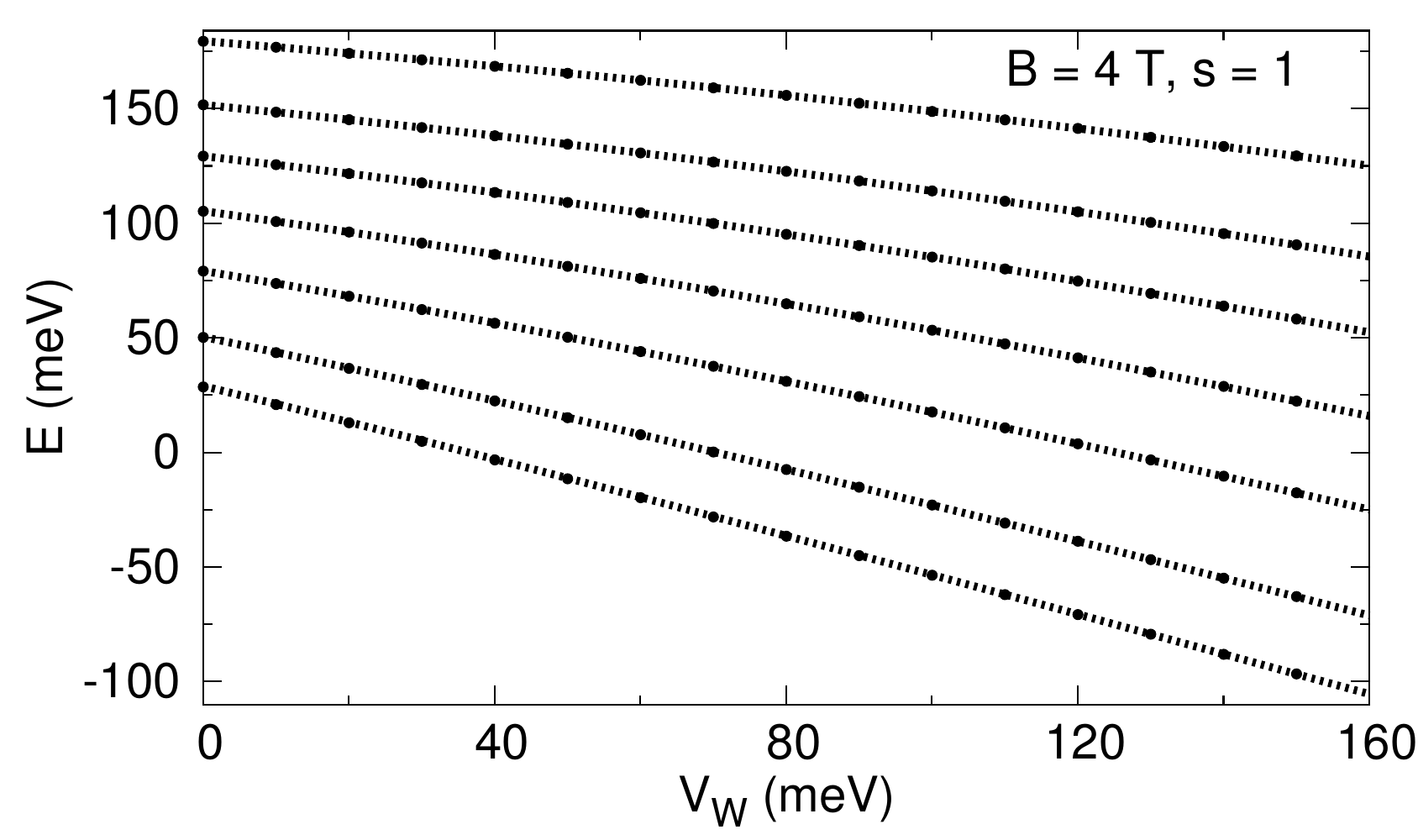}
\caption{As in Fig.~\ref{versusV} for $L_W=40$ nm and different
$B$ and $s$.}\label{ap1}
\end{figure}

\begin{figure}
\includegraphics[width=3.5cm, angle=270]{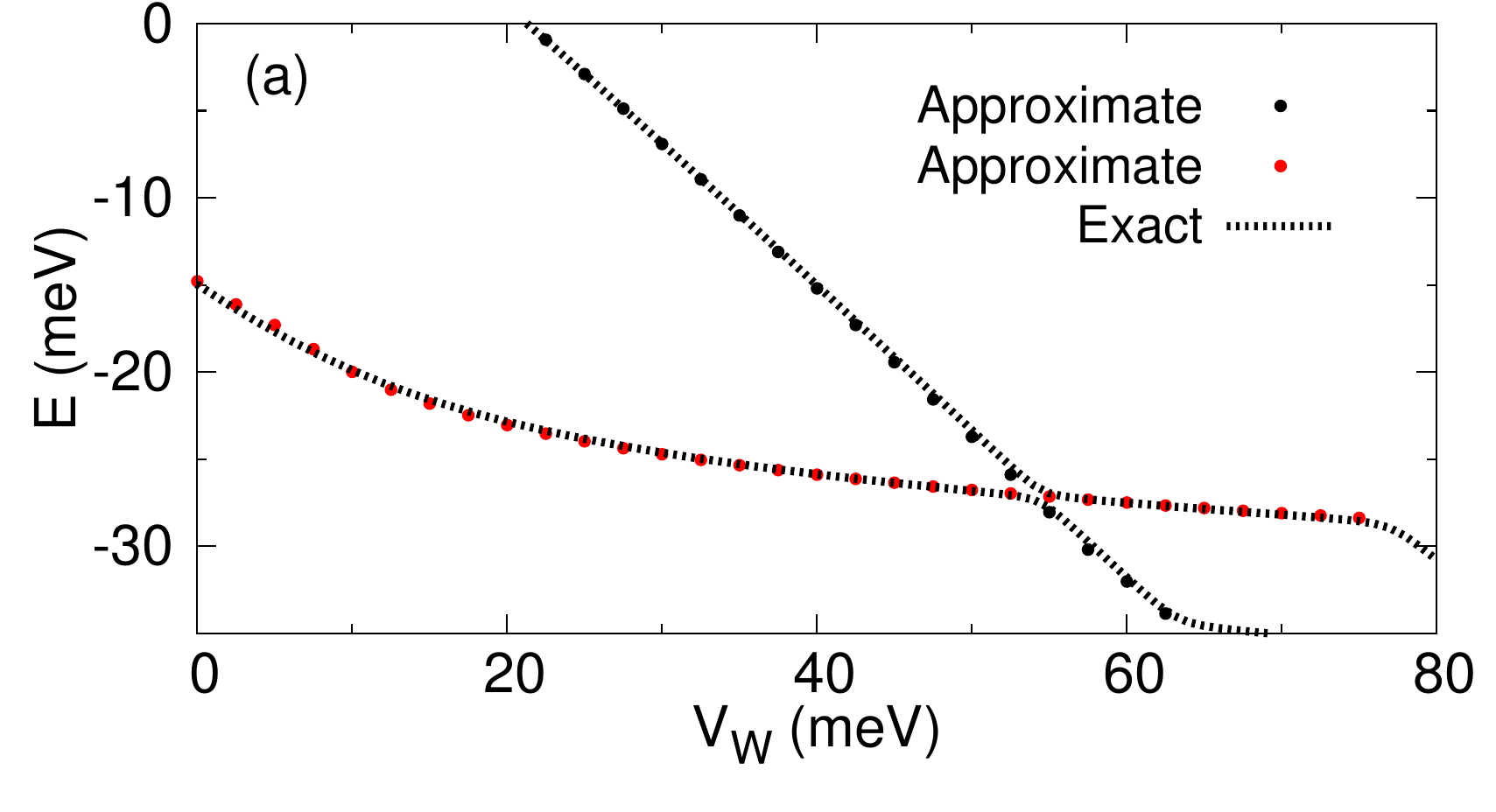}
\includegraphics[width=3.5cm, angle=270]{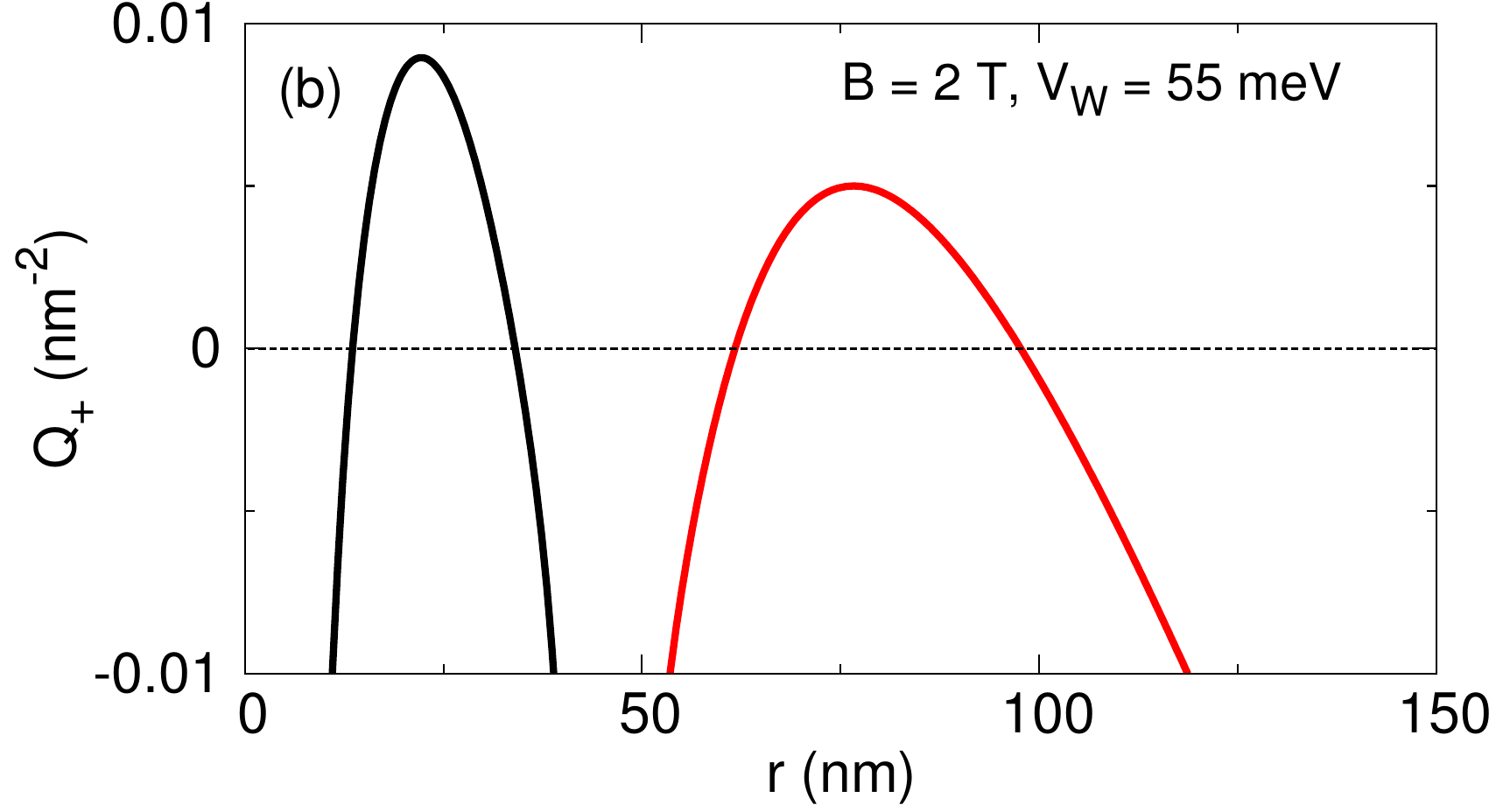}
\caption{(a) Exact and approximate QD energy levels as a function
of QD potential well for $B=2$ T, $s=2$, $m=2$. Approximate QD
levels are derived from Eq.~(\ref{oscil}) for $n=0$. (b) Near the
anticrossing point $Q_+$ has two maxima: approximate energies
shown in black (red) are derived from left (right)
maximum.}\label{ap2}
\end{figure}

\begin{figure}
\includegraphics[width=3.5cm, angle=270]{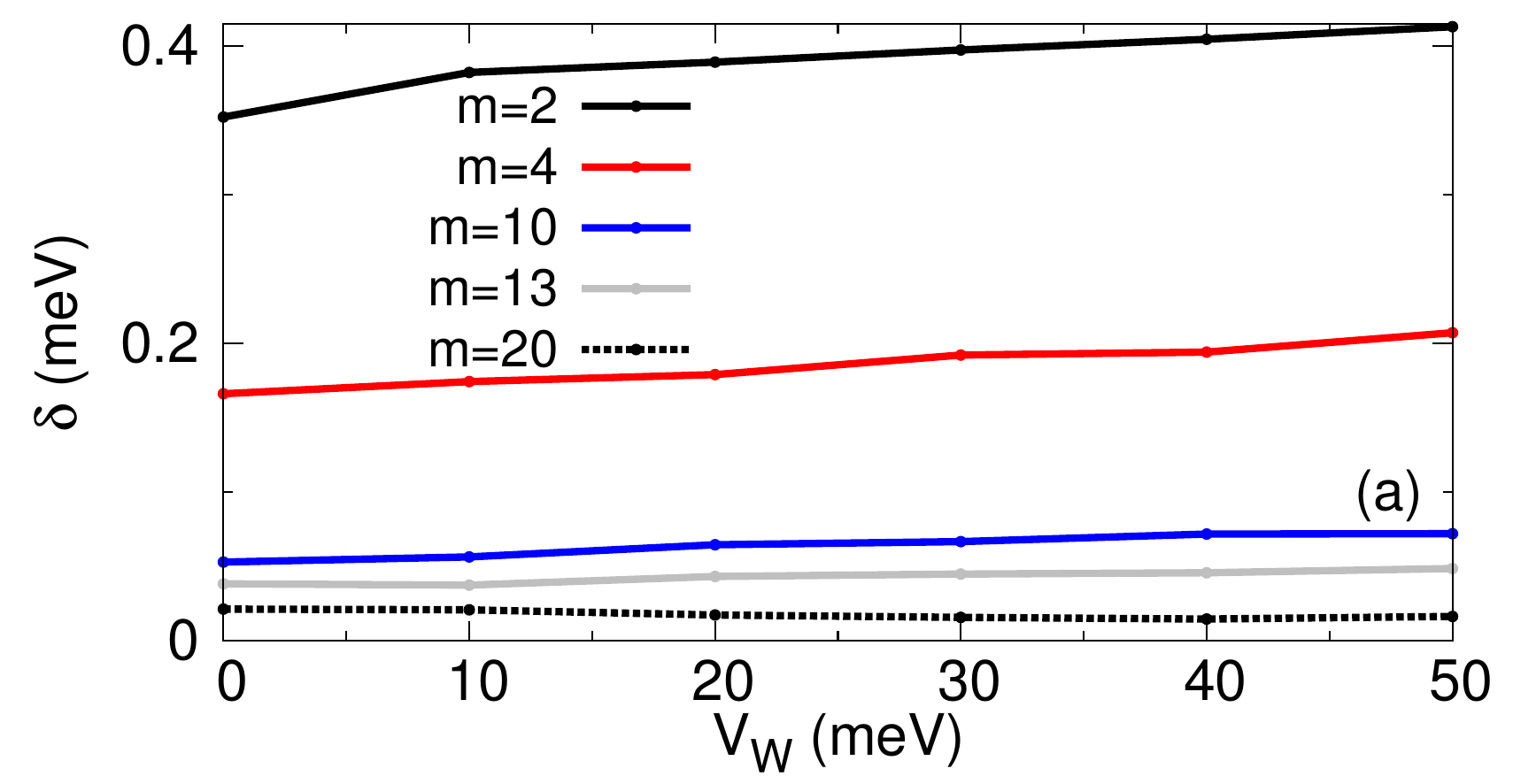}
\includegraphics[width=3.5cm, angle=270]{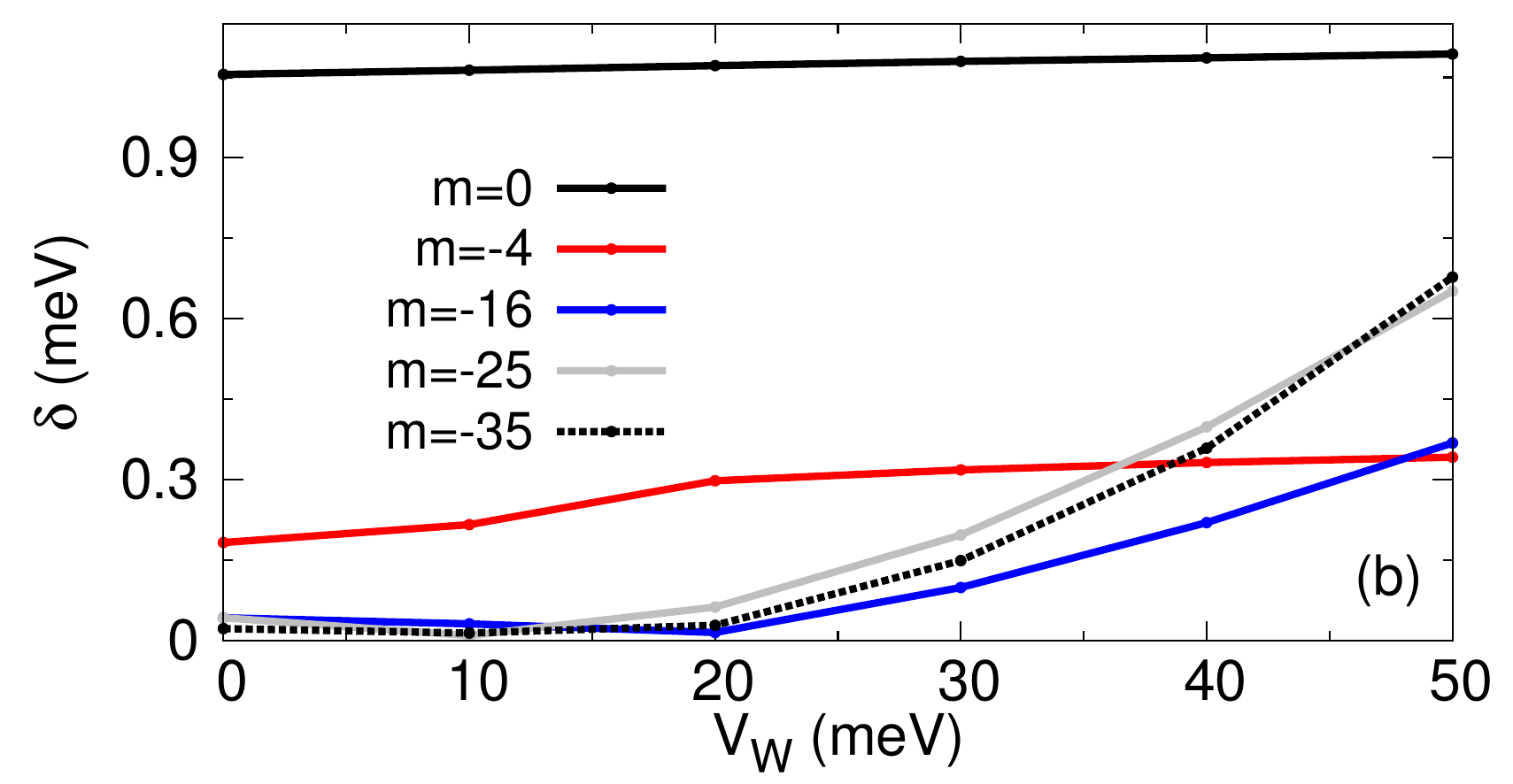}
\includegraphics[width=3.5cm, angle=270]{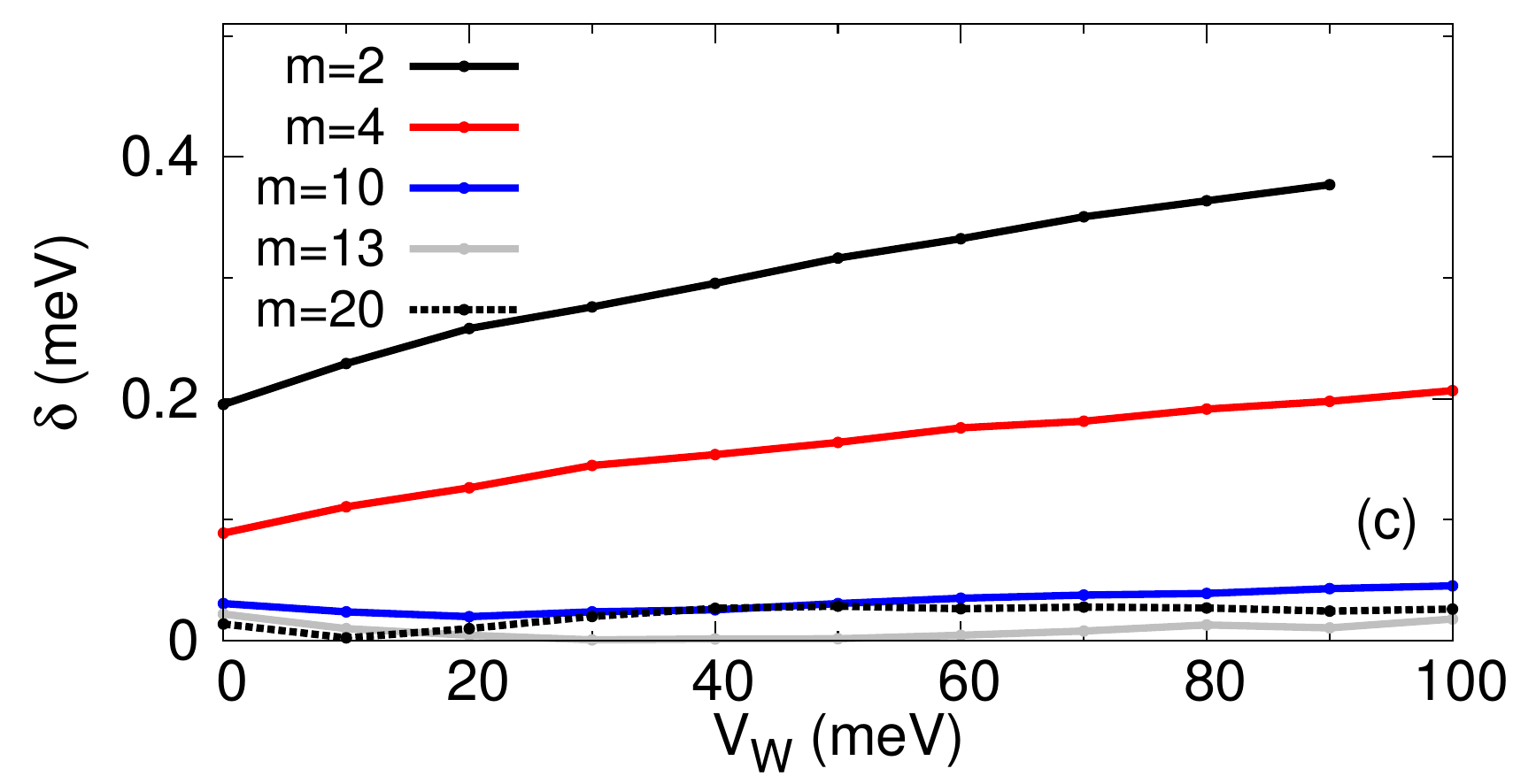}
\includegraphics[width=3.5cm, angle=270]{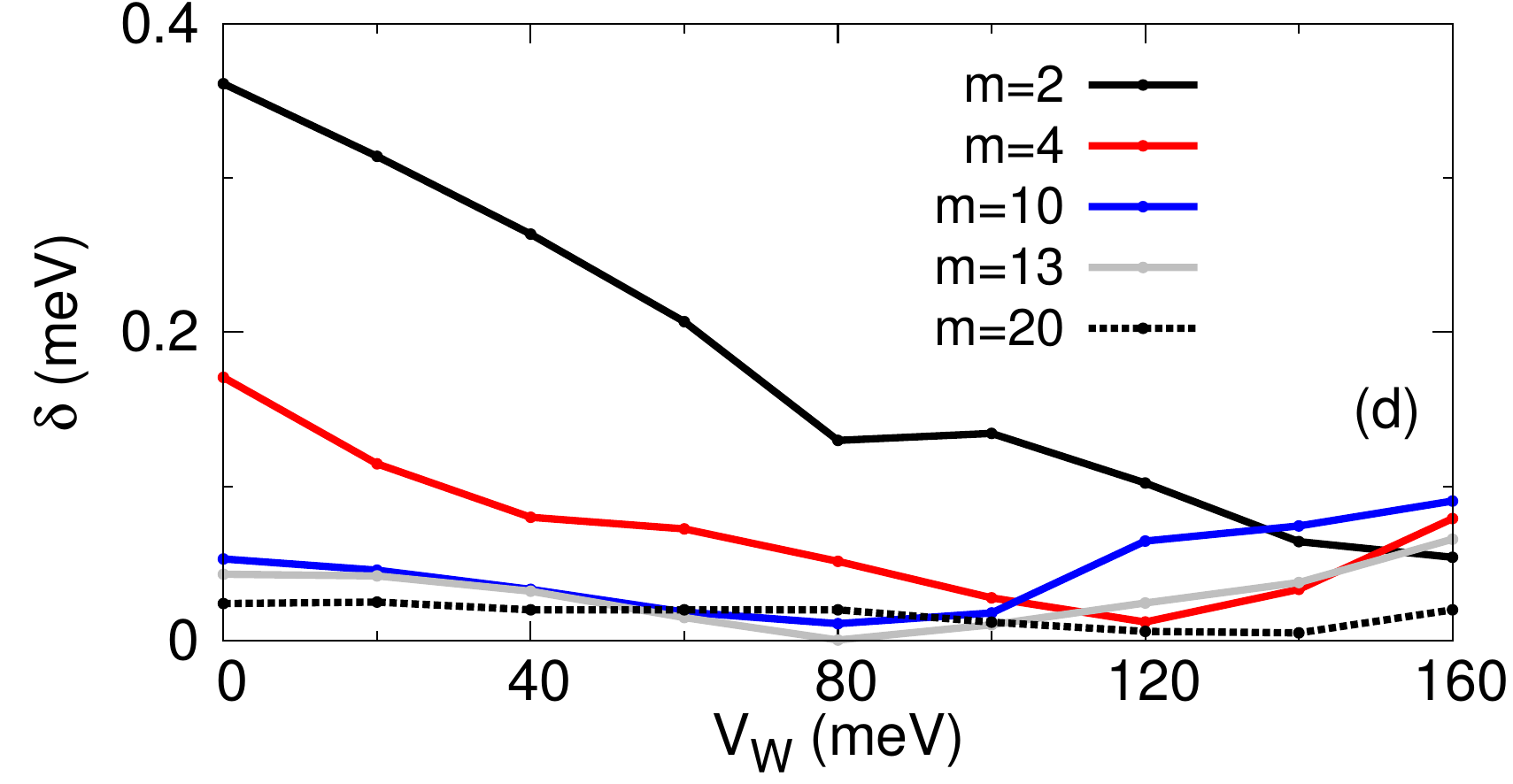}
\caption{Absolute difference between exact and approximate QD
energy levels as a function of QD potential well for different
angular momenta. Approximate QD levels are derived from
Eq.~(\ref{oscil}) for $m\ne0$ and from Eq.~(\ref{approx1}) for
$m=0$. (a) $B=4$ T, $L_W = 100$ nm, $s = 2$, (b) $B=4$ T, $L_W =
100$ nm, $s = 2$, (c) $B=2$ T, $L_W = 100$ nm, $s = 2$, and (d)
$B=4$ T, $L_W = 40$ nm, $s=1$.}\label{ap3}
\end{figure}

\appendix

\section{Examining the approximate quantum dot model}\label{wkb}

In the derivation of the approximate QD model we consider that the
QD states satisfy $w = \lambda \nu$, and examine the simplest
(adiabatic) regime where the function $\lambda$ is to a good
approximation constant. To further explore this aspect we start
with Eqs.~(\ref{l1}) and (\ref{l2}) and eliminate $\nu$ by using
Eq.~(\ref{approx1}). We derive that
\begin{equation}
\frac{1}{\lambda} = \frac{\gamma^2}{t_c \sqrt{\kappa}} (q_1 -
Q_{\pm}),
\end{equation}
and
\begin{equation}
\lambda = \frac{\gamma^2}{t_c \sqrt{\kappa}} (q_2 - Q_{\pm}),
\end{equation}
These two equations lead to
\begin{equation}\label{LAM}
 \frac{t_c \sqrt{\kappa}}{\gamma^2} \lambda^2 - (q_2-q_1) \lambda
 = \frac{t_c \sqrt{\kappa}}{\gamma^2}.
\end{equation}
If $q_1 \approx q_2$ then to a good approximation $\lambda$ is
constant, $|\lambda| \approx 1$. This limit is particularly good
for large $m$ and when the potential terms $V^{'}_{j}$,
$V^{''}_{j}$ are small. This means that the QD states should be
confined inside a region where the QD potential varies slowly.
Within a more general analysis, defining the $r$-dependent
function
\begin{equation}
\lambda_0(r) = \gamma^2 \frac{ q_2 - q_1 }{ 2 t_c \sqrt{\kappa} },
\end{equation}
and taking the limit $|\lambda_0| \ll 1$ we obtain from
Eq.~(\ref{LAM})
\begin{equation}
\lambda \approx \pm 1 + \lambda_0(r) \pm \frac{\lambda^2_0(r)}{2}.
\end{equation}
Here, we assume that any singular points can be safely ignored and
that the QD states are nicely confined within the single maximum
or two maxima of $Q_+$ as shown in Fig.~\ref{testQ}. Thus the
classical turning points, $Q_{+}=0$, define the boundaries of the
regions in which the QD states are confined. Provided $|\lambda_0|
\ll 1$ we can treat $\lambda$ as constant and the approximate QD
model is expected to be accurate.

In the approximate QD model, assuming that $w^{''}\approx \lambda
\nu^{''}$ implies that the quantity
\begin{equation}
\varepsilon = \lambda^{''} \nu + 2 \lambda^{'} \nu^{'}
\end{equation}
is vanishingly small ($\lambda$ is nearly constant) and is
therefore discarded. This process induces an error. One possible
way to estimate it is to use the state $\nu$ and energy $E$,
initially computed for $\lambda=1$, to define the ``correction''
term $\delta q_1 = \varepsilon /\lambda \nu$. This correction
comes from the exact QD equations while $\lambda$ can be computed
directly from Eq.~(\ref{LAM}). We can then replace in the
approximate Eq.~(\ref{l1}) $q_1$ with $q_1+\delta q_1$ and follow
the same steps as in the main text. The final equation has the
same form as Eq.~(\ref{approx1}) and has to be solved (exactly or
perturbatively) again for $\nu$ and $E$.

In our approach an additional error is induced due to the
expansion of $Q_{+}$ about the position of the maximum
[Eq.~(\ref{Qminus})], and the reduction of the differential
equation for $\nu$ to that of a harmonic oscillator. Therefore, to
account simultaneously for both errors we directly compare exact
and approximate energies for a few characteristic cases in
Appendix~\ref{App}. We finally remark that the oscillator can also
be used to drastically simplify $\delta q_1$ (at least) for the
nodeless state $\nu$; since $\nu^{'}/\nu = (r_0-r)Q_{+}(r_0)$ only
terms involving $\lambda$ are relevant.

\section{Additional examples of quantum dot energies}\label{App}

We continue in this appendix the investigation of the QDs for a
few more cases. Exact and approximate QD energies for different
parameters are shown in Fig.~\ref{ap1}. Similar to the results
examined in the main text (see Fig.~\ref{versusV}) the approximate
energies exhibit the correct behaviour and are in a good agreement
with the exact energies. By increasing $B$ the QD energies are
brought closer to the linear regime described by the approximate
Eq.~(\ref{linear}), while a smaller $s$ affects mostly the lowest
QD energies.

Figure~\ref{ap2}(a) presents a typical example where two QD
energies form an anticrossing point for energies $E<V_{\rm asym}$.
Here, the two QD energies emerge from a positive and a negative
Landau level respectively. As described in Sec.~\ref{cases}, the
approximate method can capture the correct dependence on the
potential well $V_W$ but the anticrossing point cannot be
resolved. Instead the approximate method predicts a crossing
point. Depending on the QD energy and the value of $V_{W}$ the
function $Q_+$ [Eq.~(\ref{qfull})] has one or two positive maxima.
Near the anticrossing point $Q_+$ has two maxima separated by a
region where a singular point dominates [Fig.~\ref{ap2}(b)]. The
two approximate QD energies in Fig.~\ref{ap2}(a) are computed by
focusing on each maximum separately, finding the corresponding
point $r_0$, and finally solving Eq.~(\ref{oscil}) for $n=0$.

Finally, in Fig.~\ref{ap3} we quantify the absolute difference,
$\delta$, between the exact and approximate energies for a few
characteristic examples. As detailed in the main text $\delta$
remains small enough enabling the approximate method to capture
all the basic characteristics of the exact QD levels derived from
the four-component model. In Fig.~\ref{ap3} the range of $V_W$
ensures that no anticrossing points are formed for $E<V_{\rm
asym}$. The dependence of $\delta$ on $V_W$ is sensitive to the
exact QD parameters and cannot be rigorously predicted, however,
we have confirmed that the same overall behaviour as in
Fig.~\ref{ap3} is found for different parameters. The worst
approximation corresponds to $m=0$ since the function $Q_{+}$ has
no positive maximum, see Sec.~\ref{cases} and Fig.~\ref{testQ}(a).
For this reason the approximate Eq.~(\ref{oscil}) cannot be
applied and for $m=0$ we need to use Eq.~(\ref{approx1}). In this
case the relative error drops below $ 5 \%$ only when $V_W \gtrsim
35$ meV. In contrast, in order to derive the approximate energies
for $m\ne0$ in Fig.~\ref{ap3} we use the harmonic oscillator
formula, Eq.~(\ref{oscil}), which offers a more pedagogical
insight. The more accurate Eq.~(\ref{approx1}) may also be used
leading to a smaller $\delta$; for example, in Fig.~\ref{ap3}(b)
the absolute difference can be reduced by one order of magnitude,
namely, $\delta\lesssim 0.07$ meV. This is a general conclusion.
In particular, if the focus is on a rigorous investigation of the
relative errors then Eq.~(\ref{approx1}) is more appropriate
especially for small QD energies near zero.

\end{document}